\documentclass[a4paper,11pt]{article}
\usepackage[utf8]{inputenc}
\usepackage[T1]{fontenc}
\usepackage[english]{babel}
\usepackage[width=160mm,height=247mm]{geometry}
\usepackage{amsmath, amsfonts, amssymb, bm}
\usepackage[pdftex]{graphicx}
\usepackage{lmodern, indentfirst, cite}
\usepackage[squaren]{SIunits}
\usepackage[textfont={footnotesize},labelfont={color=blue,bf,sf},labelsep=endash]{caption}
\usepackage[colorlinks=true,linkcolor=blue,citecolor=blue,urlcolor=blue]{hyperref}

\usepackage[affil-sl]{authblk}
\title{\textbf{\Large Scattering in Monolayer Molybdenum Disulfide Quantum Dot}}
\author[1]{Abdelhadi Belouad\thanks{\href{mailto:belabdelhadi@gmail.com}{belabdelhadi@gmail.com}}}
\author[1,2]{Abdellatif Kamal\thanks{\href{mailto:abdellatif.kamal@ensam-casa.ma}{ abdellatif.kamal@ensam-casa.ma}}}
\author[1,3]{Rachid Houça\thanks{\href{mailto:abdellatif.kamal@ensam-casa.ma}{ houca.rachid@gmail.com}}}
\author[1]{El Bouâzzaoui Choubabi\thanks{\href{mailto:abdellatif.kamal@ensam-casa.ma}{ choubabi@gmail.com}}}
\author[1]{Mohammed El Bouziani\thanks{\href{mailto:abdellatif.kamal@ensam-casa.ma}{ elbouziani@yhoo.com}}}
\affil[1]{L.P.M.C. Laboratory, Theoretical Physics Group, Faculty of Sciences, Choua\"ib Doukkali University, PO Box 20, 24000 El Jadida, Morocco}
\affil[2]{Department of Mechanical Engineering, National Higher School of Arts and Crafts, Hassan II University, Casablanca, Morocco}
\affil[3]{Equipe de Physique Théorique et Hautes Energies, Faculté des Sciences, université Ibn Zohr, PO Box 8106, Agadir, Maroc}
\date{}
\providecommand{\pacs}[1]{\noindent \textbf{PACS numbers:} #1\\}
\providecommand{\keywords}[1]{\noindent \textbf{Keywords:} #1}
\begin{document}
\begin{titlepage}
	\newgeometry{width=175mm, height=247mm}
    \maketitle
    \thispagestyle{empty}
    \vspace{3cm}
    \begin{abstract}
		We investigate the propagation of electrons in a circular quantum dot of monolayer Molybdenium disulfide $\mathrm{MoS_2}$, subjected to an electric potential. Using the continuum model, we present analytical expressions for the eigenstates, scattering coefficients, scattering efficiency, and radial component of the reflected current and electron density. We identify two scattering regimes as a function of physical parameters such as incident electronic energy, potential barrier, and quantum dot radius. For the incident electron low energy, we show that there is an appearance of scattering resonances. Also, we note that the Far-field scattered current has distinct preferred scattering directions.
	\end{abstract}
	\vspace{3cm}
\end{titlepage}
\restoregeometry
\section{Introduction}
Graphene research \cite{Novoselov05, Geim07} has stimulated the search for new two-dimensional materials. Among them, the transition-metals dichalcogenides (TMD). The monolayers of metal dichalcogenides of TMD ($\mathrm{MX_2}$ such as $\mathrm{M=Mo}$, $\mathrm{W}$; $\mathrm{X=S}$, $\mathrm{S_e}$, $\mathrm{T}$) have very recently appeared as very promising nanostructures for various applications both in the fields of optics, electronics, and spintronics. Molybdenum disulfide($\mathrm{MoS_2}$) has been a strong material known for many years, which has attracted significant attention because of its interesting electrical and optical properties \cite{Mak12, Jones13, Zaumseil14}. It is characterized by a direct band gap \cite{Mak10, Splendiani10, Tongay12, Ross13, Zeng13} in the visible frequency range and by excellent carrier mobility at room temperature \cite{Radisavljevic11, Lembke12, Lin12, Bao13, Larentis12}. This makes it a good candidate for future electronic and optoelectronic applications.

The $\mathrm{MoS_2}$ monolayer with its honeycomb atomic lattice is characterized by a strong spin-orbit interaction. This will lead to completely new electron spin properties. On the one hand, one would expect to have conduction or valence electrons that are much less sensitive to the ultrafast spin relaxation effects known for two-dimensional semiconductor structures like $\mathrm{GaAs}$ quantum wells. On the other hand, the absorption of circularly polarized light can generate a population of spin-polarized electrons (i.e. with an imbalance between the number of \emph{spin up} and \emph{spin down} electrons). Furthermore, the circularly polarized excitation makes it possible to control the distribution of these electrons in one valley or another of reciprocal space \cite{Sallen2012}. We thus speak of \emph{valley polarization}, and it is this degree of freedom that we are currently seeking to better characterize to possibly exploit it in information storage or processing applications. The $\mathrm{MoS_2}$ monolayer can be thought of as a semiconductor, the edges of the conduction and valence band are located at the two corners of the Brillouin zone, i.e. the $k$ and $k'$ points. This gives electrons and holes an additional degree of freedom which can be used for encoding information and further processing \cite{Sallen2012, Xiao07, Yao08, Gunawan06}.

Monolayer $\mathrm{MoS_2}$ quantum dots (QDs) possess distinct physical and chemical properties, strong quantum confinement properties, edge effects \cite{Wilcoxon95} and direct band gap, are important $\mathrm{MoS_2}$ based nanostructures that have attracted considerable attention from research in recent years. To date, several strategies for preparing $\mathrm{MoS_2}$ QD have been proposed, solvothermal treatment synthesis \cite{Benson15},  hydrothermal synthesis \cite{Ren15}, grinding exfoliation \cite{Xu15}, liquid exfoliation in organic solvents \cite{Damien15}, electrochemical etching \cite{Gopalakrishnan15}, and reaction processing \cite{Li14}.

The QDs exhibit edge states, which are determined by the curvatures of conduction and valence bands, localized on the edges and with energies lying in the band gap. The tight-binding model \cite{Ren15} was used to study the electronic structure of $\mathrm{MoS_2}$ quantum dots. It was shown that it is possible to  build quantum dots with the same shape but having different electronic properties due to the orbital asymmetry \cite{Ren15}.

In this work, we study the electron propagation in a circular electrostatically defined quantum dot monolayer molybdenium disulfide $\mathrm{MoS_2}$, in the presence of a potential barrier. We identify different scattering regimes depending on the radius of the quantum dot,  potential  barrier and electron
spin as well as the electron energy.

The present paper is organized as follows. In section \ref{model}, we present a theoretical study of propagation wave plane of electrons in a circular quantum dot of monolayer molybdenium disulfide $\mathrm{MoS_2}$. We give the solutions of the spinors of the Dirac equation corresponding to
each region of different scattering parameters. We use the continuity condition in order to calculate the scattering coefficients. In section \ref{Results}, we analyze the scattering efficiency, square modulus of the scattering coefficients, radial component of the far-field scattered current and electron density, we discuss our results by presenting different plot. In section \ref{Conclsion}, we present the basic conclusions of the paper.
\section{Theoretical model}\label{model}
\begin{figure}[!h]\centering
	\includegraphics[scale=0.5]{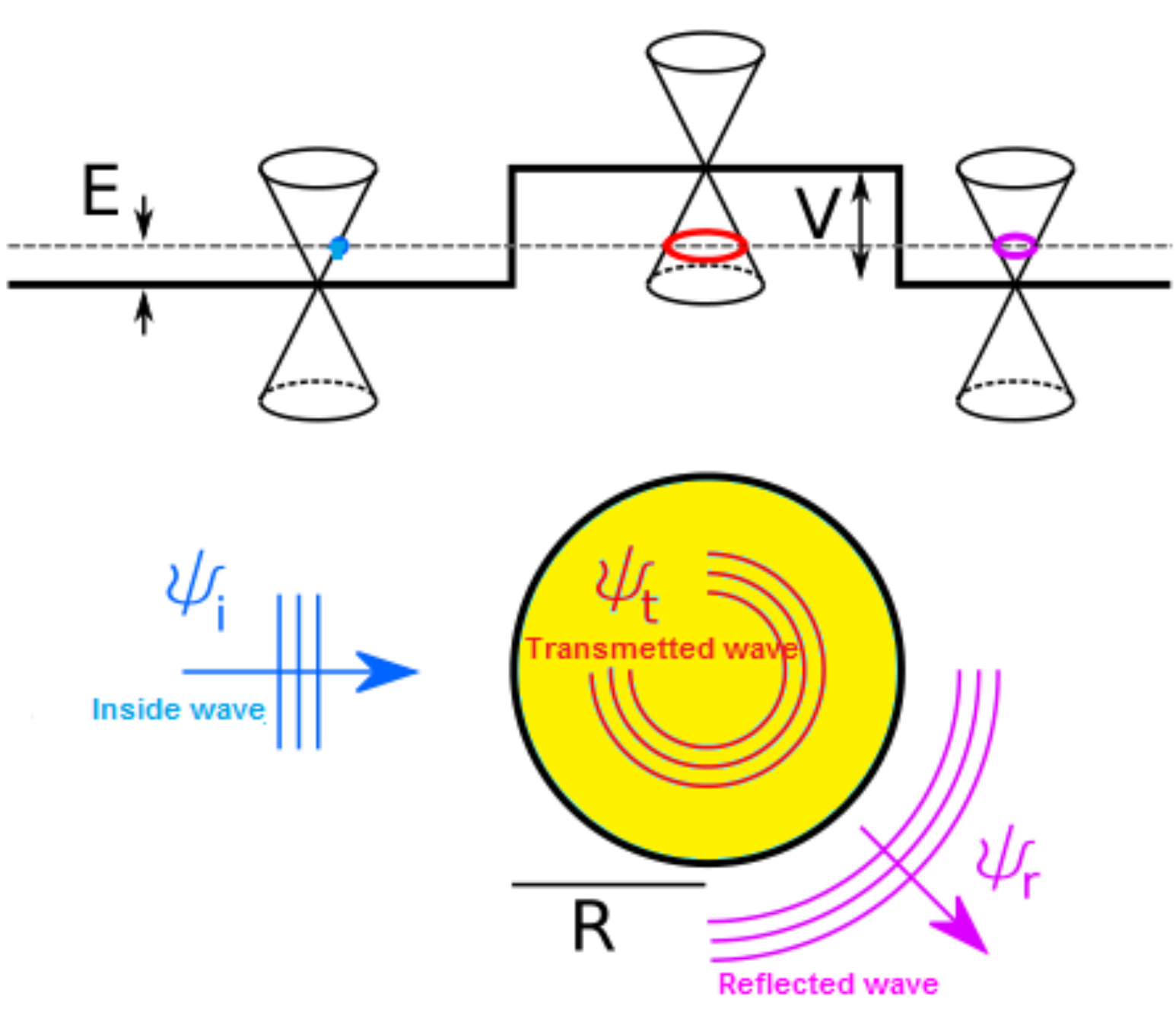}
	\caption{The energy $E$ of the Dirac electrons propagating in monolayer molybdenium disulfide $\mathrm{MoS_2}$ in a circular quantum dot. The dot is characterised by its radius $R$ and the applied bias $V$. The incident plane wave with energy $E>V$ (blue) corresponds to a state in the conduction band (upper cone). The reflected wave (purple) also lies in the conduction band, while the transmitted wave inside the dot (red) occupies the valence band (lower cone).}
	\label{f1}
\end{figure}

We consider a quantum dot of radius $R$ in the presence of a electric potentiel $V$, as illustrated in Figure \ref{f1}. In the vicinity of the $k$($\tau=1$) and  $k'$($\tau=-1$) valleys, by constructing the wave functions through the basis of conduction and valence bands, the Dirac-Weyl Hamiltonian for low-energy charge carriers in monolayer molybdenium disulfide $\mathrm{MoS_2}$ reads \cite{Oliveira16, Di Xiao12}
\begin{equation}\label{eq1}
	H=H_0+\frac{\Delta}{2} \sigma_z +\frac{\lambda_{so}}{2}\tau s_z(1-\sigma_z)+V(r)
\end{equation}
such that $H_0$ is given by
\begin{equation}\label{eq2}
	H_0= v_F\,\bm{\sigma} \cdot \bm{p}
\end{equation}
where $v_F$ is the Fermi velocity, $\bm{p}\!\!\!=\!\!\!(p_x,p_y)$ is the two-dimensional momentum
operator, $\bm{\sigma}\!=\!(\sigma_x,\sigma_y, \sigma_z)$ are Pauli matrices acting on the atomic
orbitals, V(r)  is the potential barrier, $\Delta=166$ meV \cite{Di Xiao12}  is related to the material
band gap, $s_z = \pm1$ stands for the electron \emph{spin up}  and \emph{spin down}, $\tau = \pm 1$
stands for $k$ and $k'$ valleys, $\lambda_{so}=75$ meV \cite{Di Xiao12} is the splitting of the valence
band due to spin-orbit coupling and $R$ is the dot radius. We use units such that $\hbar=1$ and the Fermi
velocity
$v_F=1$.

We now discuss localized-state solutions in our system which is modeled as a circularly symmetric quantum dot using the  potential barrier $V(r)$ and the energy gap $\Delta(r)$ respectively defined by
\begin{equation}
	V(r) =  \left\{
				\begin{array}{cc}
					0, &  r>R \\
					V, &   r\leq R
				\end{array}
			\right., \qquad
	\Delta(r) =  \left\{
					\begin{array}{cc}
						0, &  r>R \\
						\Delta, &  r\leq R
					\end{array}
				\right..
\end{equation}
We carry out our work by considering the polar coordinates (r, $\theta$), such that the Hamiltonian \eqref{eq1}  takes the form
\begin{equation}\label{eq4}
	H = \begin{pmatrix}
			V_+ & \pi^- \\
			\pi^+ & V_-+\tau s_z \lambda_{so} \\
		\end{pmatrix}
\end{equation}
where we have set the two potentials and the two operators
\begin{equation}
	V_{\pm}=V\pm\dfrac{\Delta}{2},\qquad \pi^\pm=e^{\pm i\theta}\left(-i\,\frac{\partial}{\partial r}\pm\frac{1}{r}\frac{\partial}{\partial\theta}\right).
\end{equation}
The energy spectrum is determined by solving the eigenvalue equation
\begin{equation}
	H\,\psi_m(r,\theta)=E\,\psi_m(r,\theta).
\end{equation}
Since the operator for the total angular momentum, $J_z=-i\,\hbar\,\partial_\varphi+\hbar\,\sigma_z/2$, commutes with $H$ by satisfying $[H,~J_z]=0$. This commutation requires the separability of $\psi_m$ into the radial $\Phi^{\pm}(r)$ and angular $\digamma^{\pm}(\theta)$ parts and then we have \cite{Schnez08,Heinisch13}
\begin{equation}\label{eq5}
	\psi_m(r,\theta) = \Phi_{m}^{+}(r)\,\digamma_{m}^{+}(\theta) + \Phi_{m+1}^{-}(r)\,\digamma_{m+1}^{-}(\theta)
\end{equation}
where the two angular components are
\begin{equation}\label{eq6}
	\digamma_m^+(\theta)=\frac{e^{im\theta}}{\sqrt{2\pi}}\begin{pmatrix}
															1\\
															0\\
														\end{pmatrix},\qquad
	\digamma_{m+1}^-(\theta)=\frac{e^{i(m+1)\theta}}{\sqrt{2\pi}}\begin{pmatrix}
																	0\\
																	1\\
																\end{pmatrix}
\end{equation}
and $m=0,~ \pm 1,~ \pm 2,~ \cdots$, is the total angular quantum number.

In order to get the solutions of the energy spectrum, we have to solve the eigenvalue problem
\begin{equation}
	H\,\psi_m(r,\theta)=E\,\psi_m(r,\theta)
\end{equation}
by considering two regions according to Figure \ref{f1} : outside ($r>R$) and inside ($r\leq R$) the quantum dot. Thus, we have an incident wave $\psi_i$ propagation in the $x$-direction, the reflected wave $\psi_r$ is an outgoing wave and a transmitted wave $\psi_t$ inside the dot.

Outside the dot ($r>R$), the radial components $\Phi_{m}^{+}(r)$ and $\Phi_{m+1}^{-}(r)$ satisfy two coupled differential equations
\begin{eqnarray}
	&&-i\,\frac{\partial}{\partial r}\Phi_{m}^{+}(r)+i\,\frac{m}{r}\,\Phi_{m}^{+}(r)=E\,\Phi_{m+1}^{-}(r)\label{eq7}\\
	&& -i\,\frac{\partial}{\partial r}\Phi_{m+1}^{-}(r)-i\,\frac{m+1}{r}\,\Phi_{m+1}^{-}(r)=E\,\Phi_{m}^{+}(r).\label{eq8}
\end{eqnarray}
which can be handled by injecting \eqref{eq7} into \eqref{eq8} to derive a second differential equation satisfied by $\Phi_{m}^{+}(r)$
\begin{equation}\label{eq9}
	\left(r^2\frac{\partial^2}{\partial^2 r}+r\frac{\partial}{\partial r}+r^2E^2-m^2\right)\Phi_{m}^{+}(r)=0
\end{equation}
where the solutions are the Bessel functions $J_{m}(Er)$. Moreover, the wave function of the incident electron, propagating along $x$-direction ($x=r\cos\theta$), takes the form
\begin{eqnarray}\label{eq11}
	\psi^i_m(r,\theta) &=& \frac{1}{\sqrt{2}}\sum_m i^m\,
	\left[
		J_{m}(kr)\,e^{im\theta}
		\begin{pmatrix}
			1\\
			0\\
		\end{pmatrix}
		+i\,J_{m+1}(kr)\,e^{i(m+1)\theta}
		\begin{pmatrix}
			0\\
			1\\
		\end{pmatrix}
	\right]
\end{eqnarray}
as well as the reflected wave
\begin{equation}\label{eq12}
	\psi^r_m(r,\theta) = \frac{1}{\sqrt{2}}\sum_m i^m\,\alpha_{m}
	\left[
		H^{(1)}_{m}(kr)\,e^{im\theta}\begin{pmatrix}
												1\\
												0\\
											\end{pmatrix}
		+i\,H^{(1)}_{m+1}(kr)\,e^{i(m+1)\theta}\begin{pmatrix}
												0\\
												1\\
											\end{pmatrix}
	\right]
\end{equation}
wehere  $H^{(1)}_{m}(kr)$ is the Hankel functions of the first kind \cite{Berry13} and $\alpha_{m}$ is the scattering coefficients and the wave number $k=E$.

Inside the dot ($r\leq R$), we obtain the following equations corresponding to the radial functions $\Phi_{m}^{+}$ and $\Phi_{m+1}^{-}$
\begin{eqnarray}
	&& i\left(\frac{\partial}{\partial r}-\frac{m}{r}\right)\Phi_{m}^{+}(r)+\left(E-V_--\tau s_z \lambda_{so}\right)\Phi_{m+1}^{-}(r)=0 \label{eq13}\\
	&& i\left(\frac{\partial}{\partial r}+\frac{m+1}{r}\right)\Phi_{m+1}^{-}(r)+\left(E-V_+\right)\Phi_{m}^{+}(r)=0 \label{eq14}
\end{eqnarray}
Expressing \eqref{eq13} as
\begin{equation}\label{eq14e}
	\Phi_{m+1}^{-}(r) =  -\frac{i}{E-V_--\tau s_z \lambda_{so}}\left(\frac{\partial}{\partial r}-\frac{m}{r}\right)\Phi_{m}^{+}(r)
\end{equation}
and replacing it in \eqref{eq14} we get a differential equation for $\Phi_{m}^{+}(r)$
\begin{equation}\label{eq15}
	\left(r^2\frac{\partial^2}{\partial^2 r}+r\frac{\partial}{\partial r}+r^2\gamma^2-m^2 \right)\Phi_{m}^{+}(r)=0
\end{equation}
where
\begin{equation}
	\gamma^2=\left(E-V_+\right)(E-V_--\tau s_z \lambda_{so}).
\end{equation}
The solution of \eqref{eq15} can be worked out to get the transmitted wave as
\begin{equation}\label{eq16}
	\psi^t_m(r,\theta) = \frac{1}{\sqrt{2}}\,\sum_m i^m\,\beta_{m}
	\left[
		J_{m}(\gamma r)\,e^{im\theta}
		\begin{pmatrix}
			1\\
			0\\
		\end{pmatrix}
		+i\,\mu\,J_{m+1}(\gamma r)\,e^{i(m+1)\theta}
		\begin{pmatrix}
			0\\
			1\\
		\end{pmatrix}
	\right]
\end{equation}
where the $\beta_m$ denote the transmission coefficients and
\begin{equation}
	\mu= \sqrt{\frac{E-V_+}{E-V_--\tau s_z \lambda_{so}}}.
\end{equation}
Requiring the eigenspinors continuity at the boundary $r=R$ of the quantum dot,
\begin{equation}
	\psi^i_m(R)+\psi^r_m(R)=\psi^t_m(R),
\end{equation}
to obtain the conditions
\begin{eqnarray}
	&& J_{m}(kR)+\alpha_{m}\,H^{(1)}(kR)=\beta_{m}\,J_{m}(\gamma R), \label{eq17}\\
	&& J_{m+1}(kR)+\alpha_{m}\,H_{m+1}^{(1)}(kR)=\mu\, \beta_{m}\,J_{m+1}(\gamma R). \label{eq18}
\end{eqnarray}
Solving these equations to get the scattering coefficients
\begin {equation}\label{eq19}
	\alpha_{m}=-\frac{J_{m}(\gamma R)\,J_{m+1}(kR)-\mu\, J_{m+1}(\gamma R)\, J_{m}(kR)}{J_{m}(\gamma R)\,H^{(1)}_{m+1}(kR)-\mu\, J_{m+1}(\gamma R)\, H^{(1)}_{m}(kR)}
\end {equation}
and the transmission coefficients by
\begin {equation}\label{eq20}
	\beta_{m}=\frac{J_{m}(kR)\,H^{(1)}_{m+1}(kR)-J_{m+1}(kR)\,H_{m}^{(1)}(kR)}{J_{m}(\gamma R)\,H^{(1)}_{m+1}(kR)-\mu\, J_{m+1}(\gamma R)\, H^{(1)}_{m}(kR)}
\end {equation}
The current density is defined by ${\bm j}=\psi^{\dag}{\bm\sigma} \psi$ where inside the dot we have $\psi=\psi_t$, however, outside the dot we have $\psi=\psi_i+\psi_r $.

The far-field radial component of the reflected current, which characterizes angular scattering, reads
\begin{equation}
	j_{r}= \psi^{\dag}\left(\sigma_{x} \cos\theta+\sigma_{y}\sin\theta\right)\psi=\psi^{\dag}
	\begin{pmatrix}
		0 & e^{-i\theta} \\
		e^{i\theta} & 0 \\
	\end{pmatrix}
	\psi
\end{equation}
the corresponding radial current can be written as
\begin{eqnarray}\label{eq23e}
	j^{r}_{r}=\frac{1}{2}\sum^{\infty}_{m=0}A_m(kr) \times
	\begin{pmatrix}
		0 & e^{-i\theta} \\
		e^{i\theta} & 0 \\
	\end{pmatrix}
	\times \sum^{\infty}_{m=0}A_m^*(kr)
\end{eqnarray}
where
\begin{equation}\label{eq24e}
	A_{m}(kr) = (-i)^m\left[
								H_m^{(1)*}(kr)
								\begin{pmatrix}
									\alpha_m^*\,e^{-im\theta} \\
									\alpha^*_{-m-1}\,e^{im\theta}
								\end{pmatrix}
								-i\, H_{m+1}^{(1)*}(kr)
								\begin{pmatrix}
									\alpha^*_{-m-1}\,e^{i(m+1)\theta} \\
									\alpha_m^*\, e^{-(m+1)\theta}
								\end{pmatrix}
							\right]
\end{equation}
By injecting the asymptotic behavior of the Hankel function of the first kind for $kr\gg 1$
\begin{eqnarray}\label{eq26e}
	\displaystyle H_m(kr)\simeq \sqrt{\dfrac{2}{\pi kr}}\,e^{i\left(kr-\frac{m\pi}{2}-\frac{\pi}{4}\right)},
\end{eqnarray}
into \eqref{eq24e}, the current density \eqref{eq23e} can be reduced to the following
\begin{equation}\label{eq24}
	j^{r}_{r}(\theta)=\frac{4}{\pi kr}\sum^{\infty}_{m=0}\left[1+\cos\left(2m+1\right)\theta\right]\,|c_m|^2
\end{equation}
where
\begin{equation}
	|c_m|^2 = \frac{1}{2}\left[{|\alpha_m|^2 + |\alpha_{-(m+1)}|^2}\right].
\end{equation}
The scattering cross section $\sigma$ is defined by \cite{Grujic11}
\begin{equation}\label{eq28}
\displaystyle\sigma=\dfrac{I_r^r}{I^i/A_u}
\end{equation}
where $I_r^r$ is the total reflected flux through a concentric circle and $(I^i/A_u)$ is the incident flux per unit area. Moreover, $I_r^r$ is given by
\begin{equation}
	I^{r}_{r}=\int_0^{2\pi} j^{r}_{r}(\theta)\,r\,\mathrm{d}\theta=\frac{8}{k}\sum^{\infty}_{m=0}|c_m|^2.
\end{equation}

We note that for the incident wave \eqref{eq11}, we have $I^i/A_u=1$.

To go deeply in our study for the scattering problem for a plane Dirac electron for different size of the circular quantum dot, we define the scattering efficiency $Q$ by dividing the scattering cross section by the geometric cross section. It is given by
 \begin{equation}
Q=\frac{\sigma}{2R}=\frac{4}{kR}\sum^{\infty}_{m=0}|c_m|^2.
\end{equation}
\section{Results and discussions}\label{Results}
\begin{figure}[!h]\centering
	\includegraphics[scale=0.6]{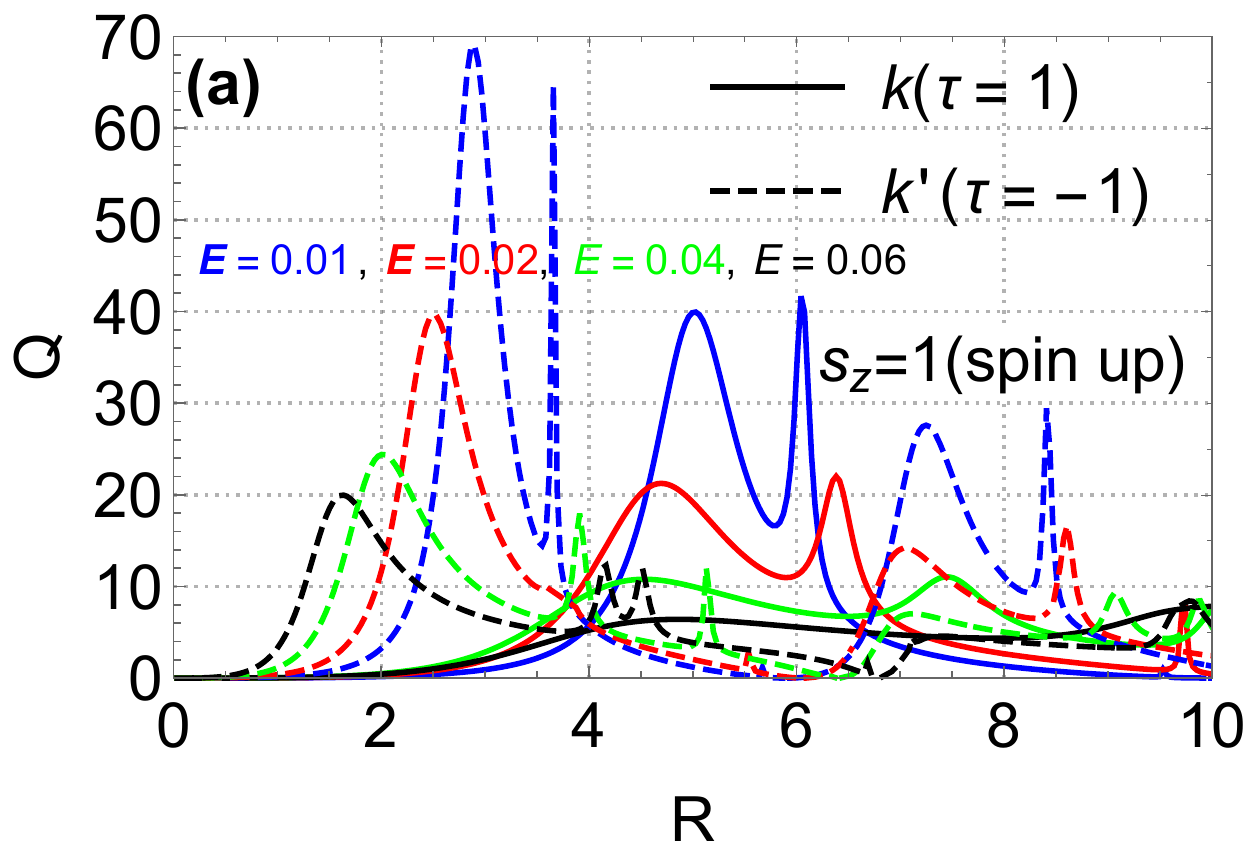}\hfill \includegraphics[scale=0.6]{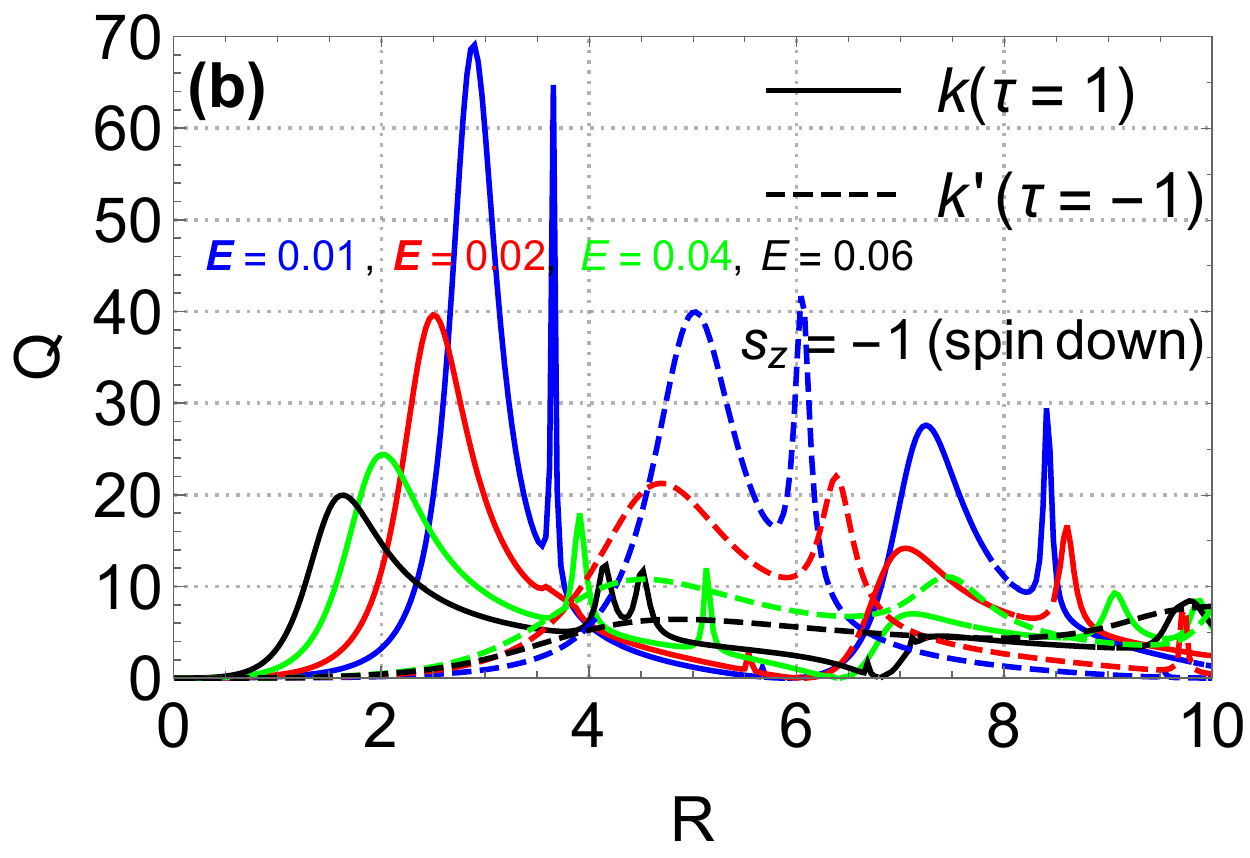}\\
	\includegraphics[scale=0.6]{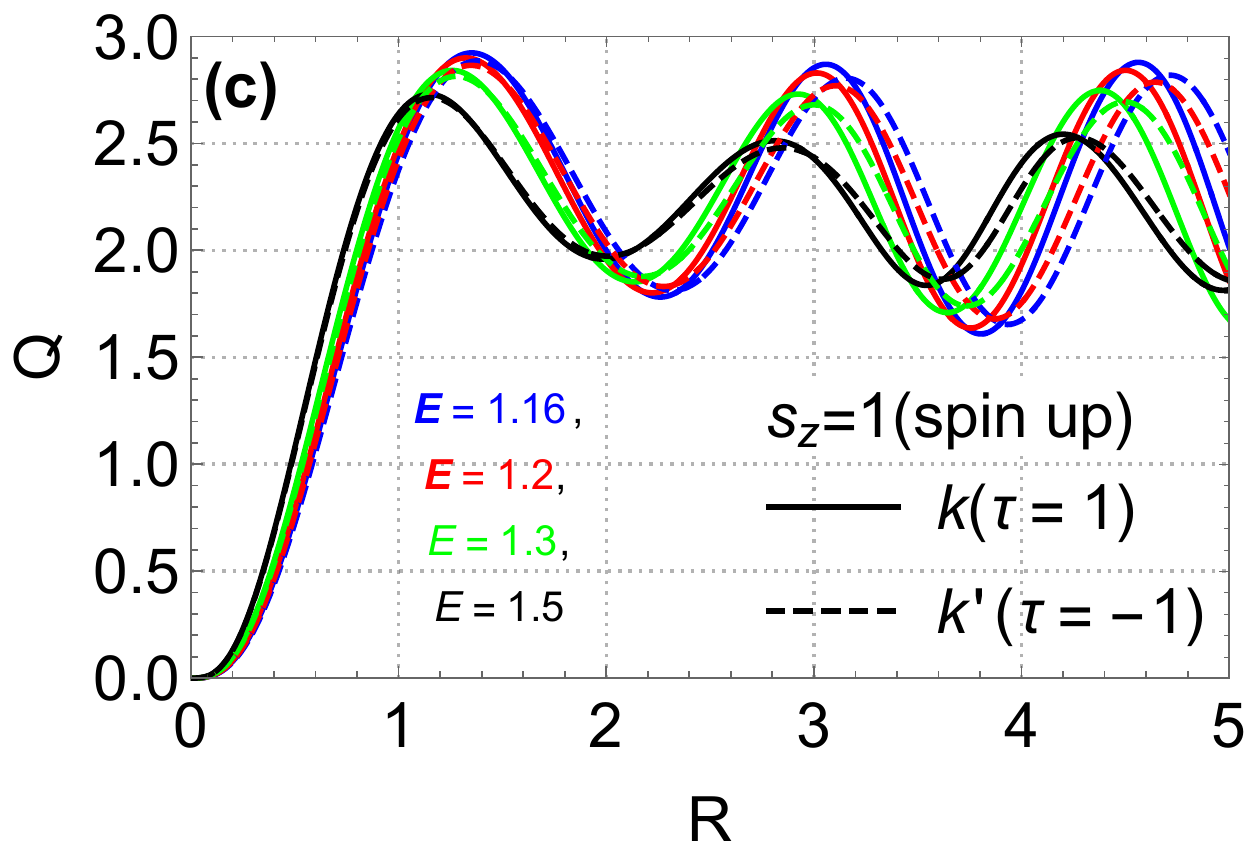}\hfill \includegraphics[scale=0.6]{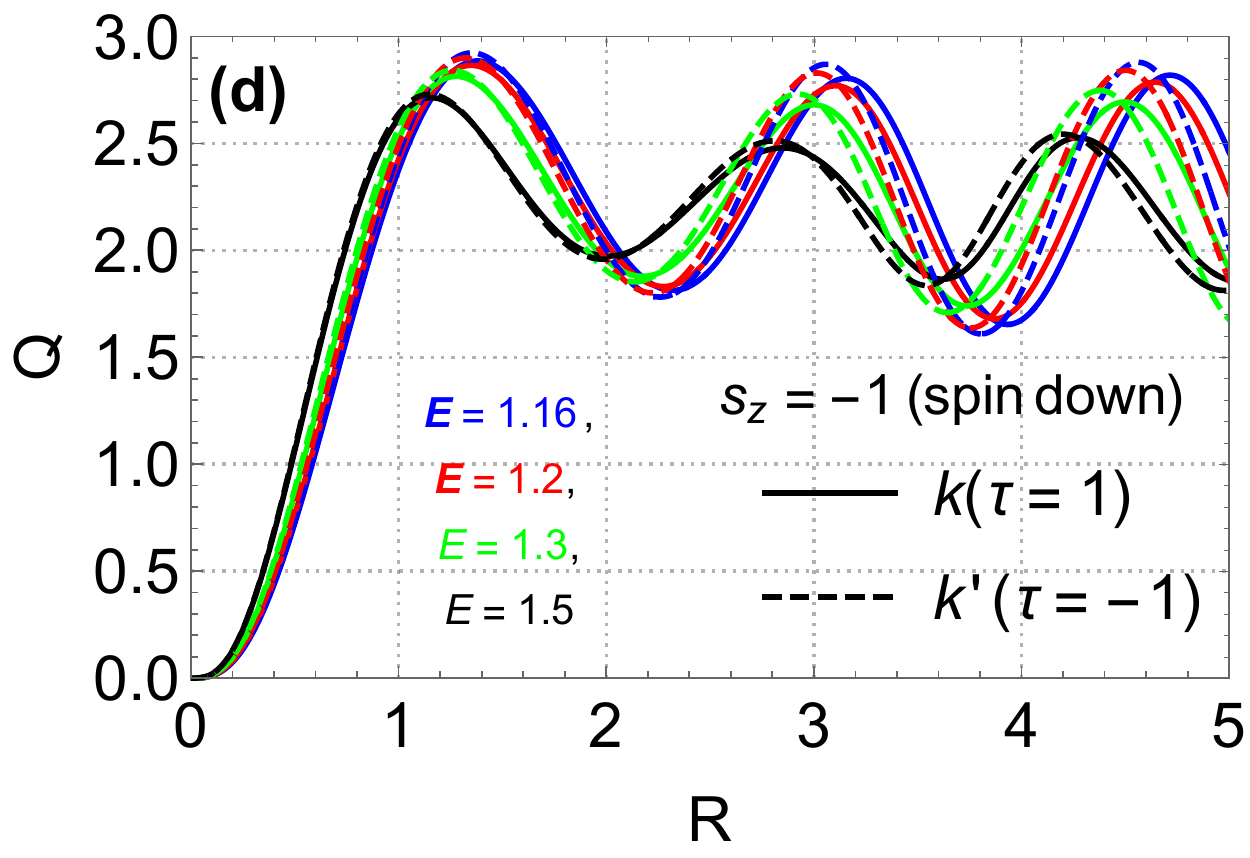}
	\caption{Scattering efficiency $Q$, for the potential $V = 1$, as a function of the dot radius $R$ at $E=0.01,~ 0.02,~ 0.04,~ 0.06$ for (a): the \emph{spin up} state in two valleys $k$($\tau=1$, $s_z=1$) and $k'$($\tau=-1$, $s_z=1$) and (b): the \emph{spin down} state in two valleys $k$($\tau=1$, $s_z=-1$) and $k'$($\tau=-1$, $s_z=-1$) and at $E=1.16,~ 1.2,~ 1.3,~ 1.5$ for (c): the \emph{spin up} state in two valleys k($\tau=1$, $s_z=1$) and $k'$($\tau=-1$, $s_z=1$) and (d): the \emph{spin down} state in two valleys $k$($\tau=1$, $s_z=-1$) and $k'$($\tau=-1$, $s_z=-1$).}
	\label{Fig2}
\end{figure}

In Figures \ref{Fig2}(a) and \ref{Fig2}(b), we show the scattering efficiency $Q$ for low energies under the barrier scattering ($n$-$p$ junction) $E = 0.01,~0.02,~0.04,~0.04 <V=1$, as a function of radius of quantum dot $R$ for the \emph{spin up} state in the two valleys $k$($\tau=1$, $s_z=1$) and $k'$($\tau=-1$, $s_z=1$) and the \emph{spin down} state in the two valleys $k$($\tau=1$, $s_z=-1$) and $k'$($\tau=-1$, $s_z=-1$). We notice that as $R\rightarrow 0$, $Q\rightarrow 0$, when $R$ increases, $Q$ increases to a maximum value $Q_{max}=20,~25,~39.5,~68$ for $E = 0.06,~0.04,~0.02,~0.01$ respectively, for the state ($k'$, $s_z$) (Figure \ref{Fig2}(a)) and for the state ($k$, -$s_z$) (Figure \ref{Fig2}(b)), then the scattering efficiency $Q$ has a strongly damped oscillatory behavior with the appearance of net transverse resonant peaks, similar to graphene quantum dots \cite{Grujic11,Schulz15, Belouad18}. As the radius $R$ increases, the height of the peak is reduced but its width becomes larger which shows the peculiarity of the energy dispersion becomes apparent. Moreover, by comparing Figures \ref{Fig2}(a) and \ref{Fig2}(b), we find that the dependence of $Q$ for the \emph{spin up} and \emph{spin down} states in the two valleys is symmetrical i.e. $Q(-\tau, s_z)=Q(\tau, -s_z)$.

Figures \ref{Fig2}(c) and \ref{Fig2}(d) show, respectively, the scattering efficiency $Q$ for energies over the barrier ($n$-$n$ junction) $E = 1.16,~1.2,~ 1.3,~ 1.5 >V=1$,  as a function of radius of quantum dot $R$ for the \emph{spin up} state in the two valleys $k$($\tau=1$, $s_z=1$) and $k'$($\tau=-1$, $s_z=1$) and the \emph{spin down} state in the two valleys $k$($\tau=1$, $s_z=-1$) and $k'$($\tau=-1$, $s_z=1$). Furthermore, by increasing $R$, the scattering efficiency $Q$ increases almost linearly until a maximum value $Q_{max}=2.9$ corresponds to a specific value of $R$. However, by increasing $R$ and for the four values of energy, the four curves are exhibits oscillatory behavior \cite{Zheng19}. In this regime ($E>V$), we find behaviors of symmetry $Q$ ($Q(-\tau, s_z)=Q(\tau, -s_z)$) with respect to $s_z$ and $\tau$ similar to those of the preceding regime ($E<V$).
\begin{figure}[!h]\centering
	\includegraphics[scale=0.4]{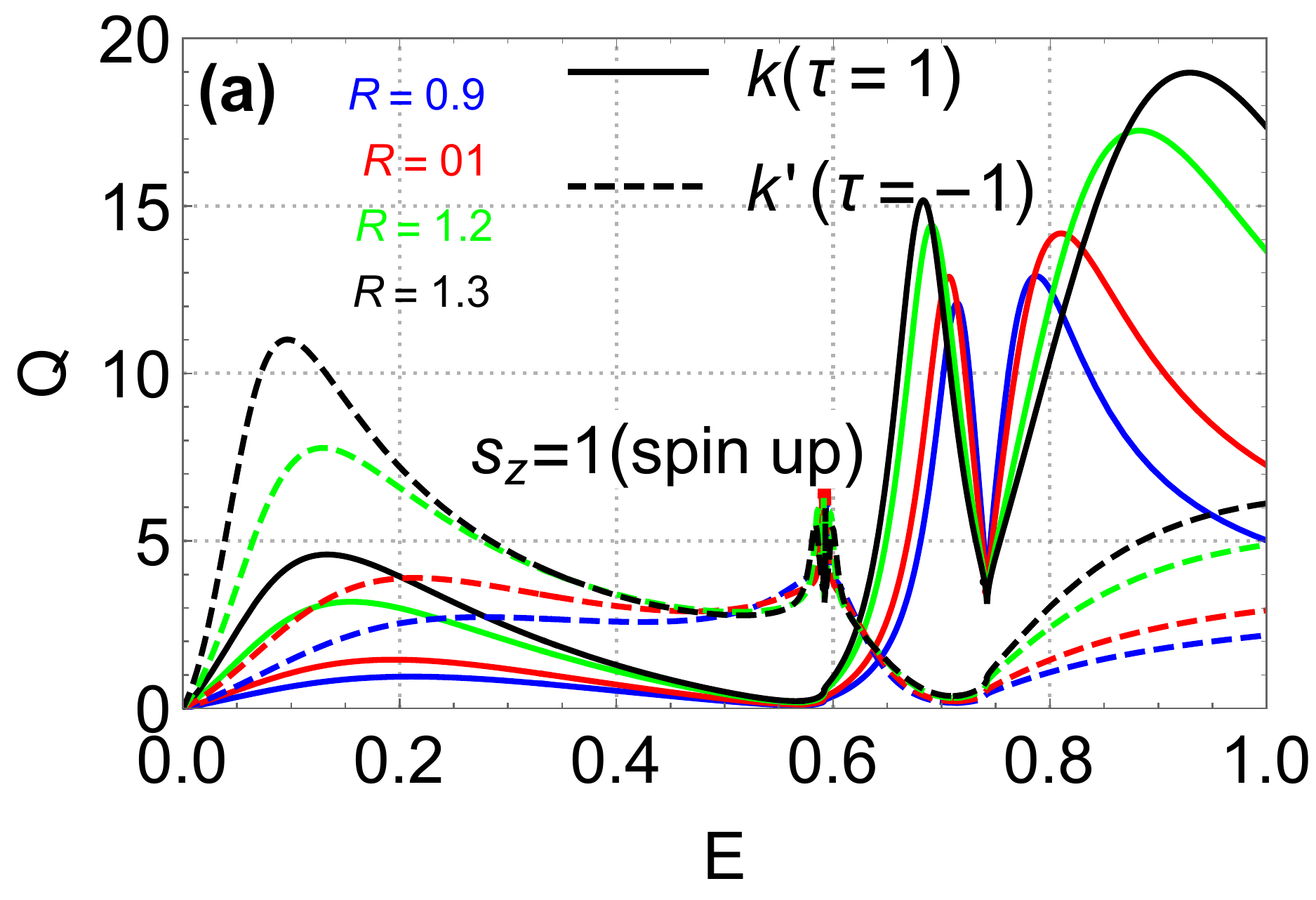}\hspace{0.5cm}\includegraphics[scale=0.4]{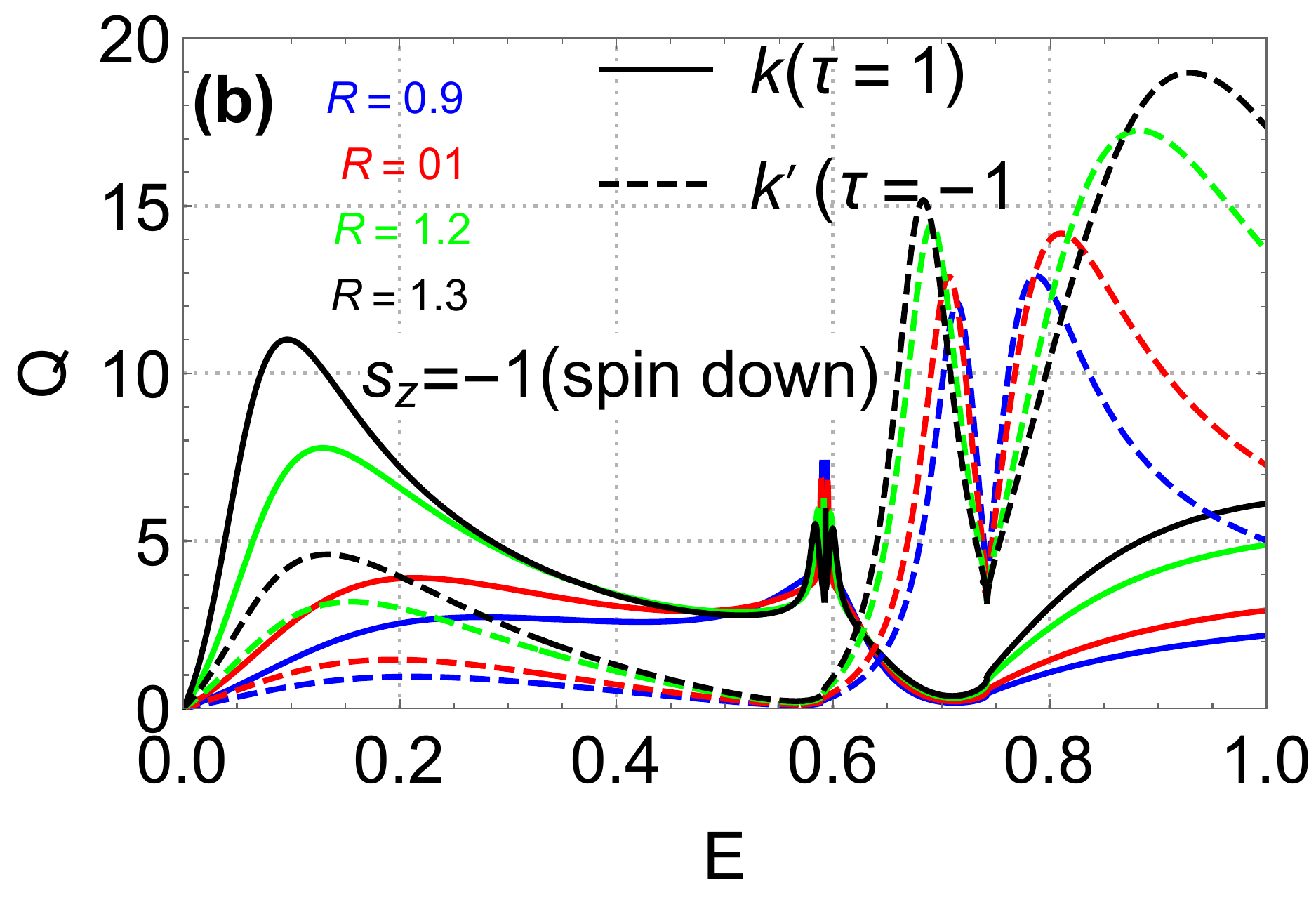}
	\includegraphics[scale=0.4]{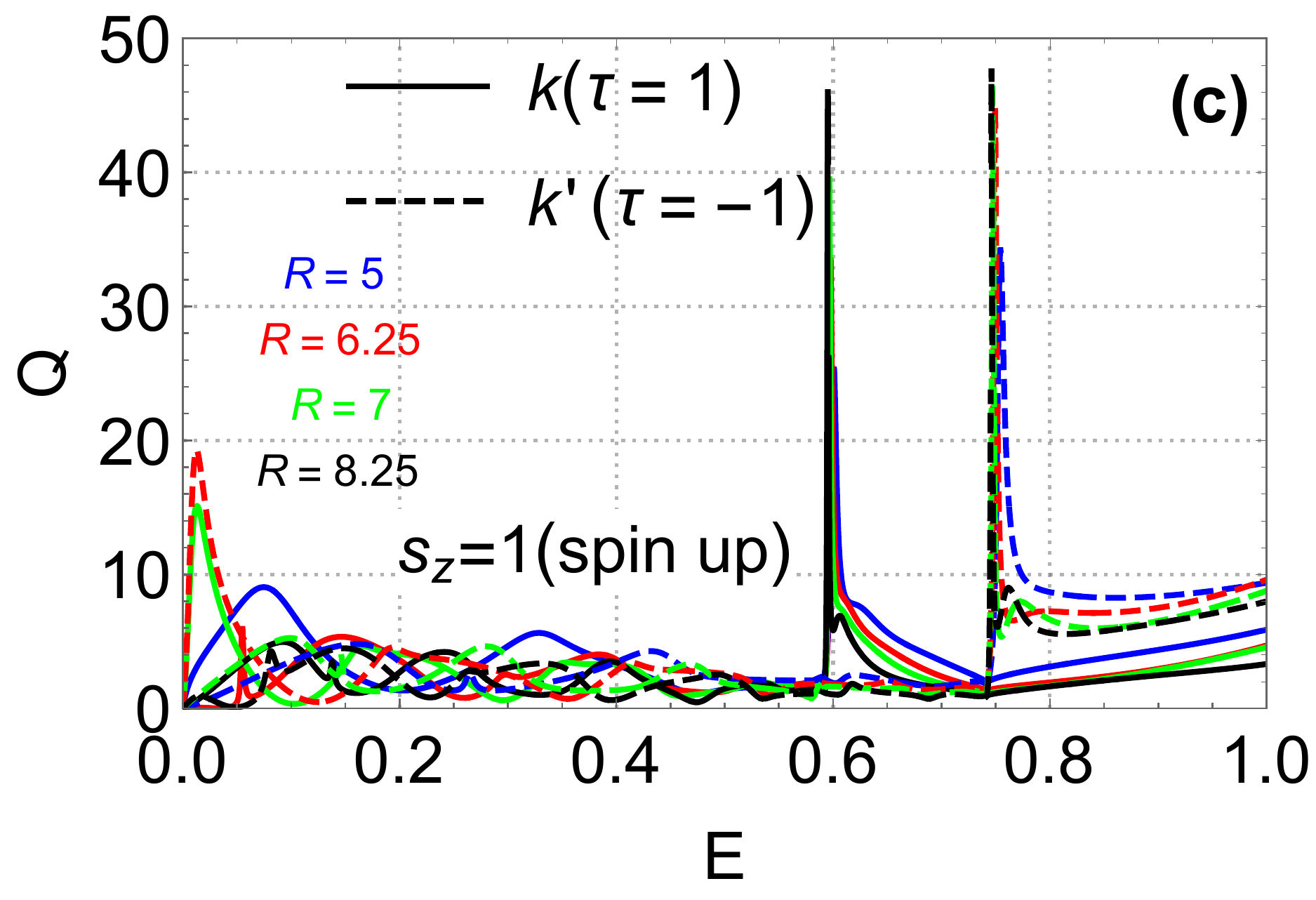}\hspace{0.5cm}\includegraphics[scale=0.4]{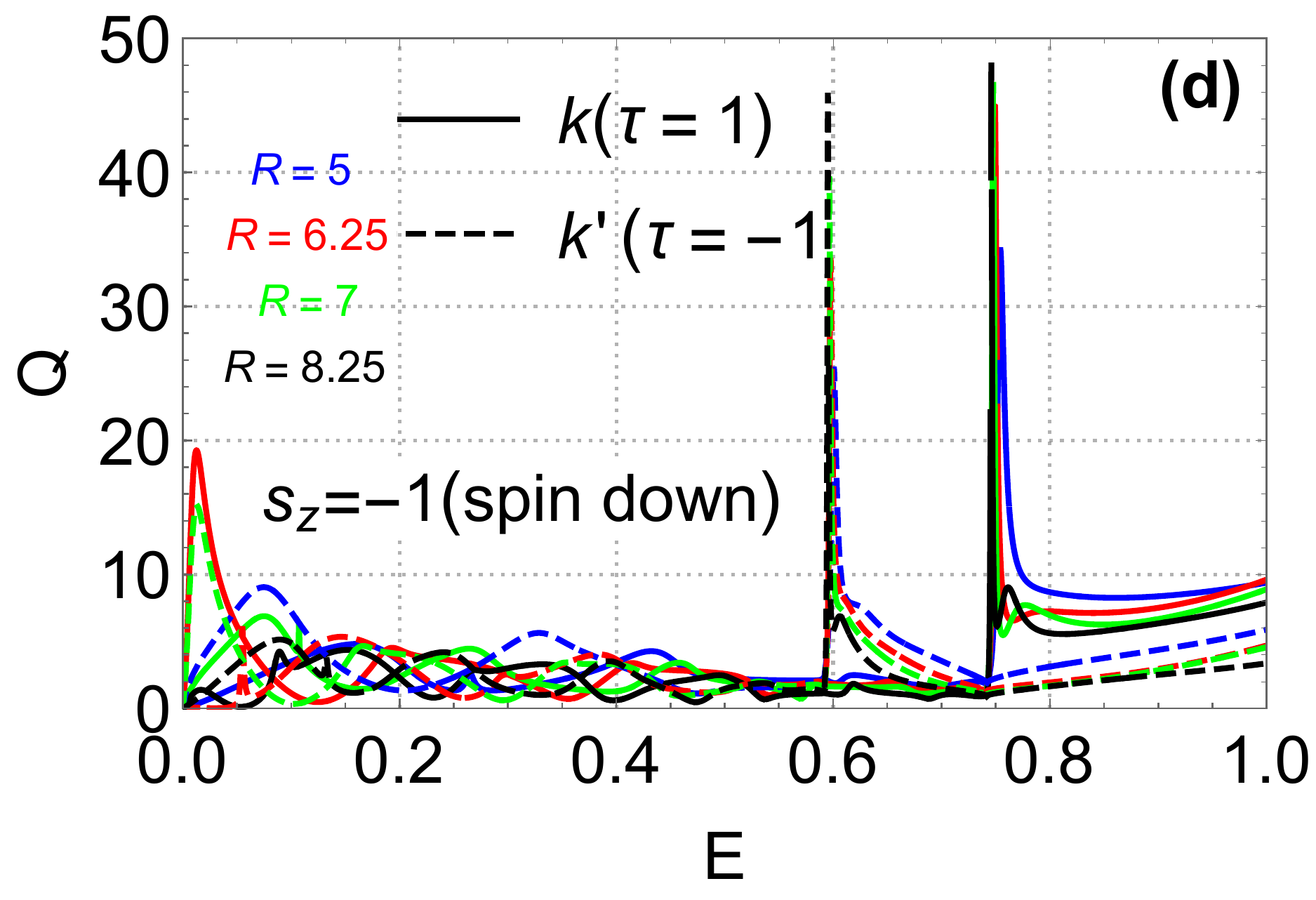}
	\caption{Scattering efficiency $Q$, for the potential $V=1$, as a function of the energy E of the incident electron  at $R=0.9,~1,~ 1.2,~ 1.3$ for (a): the \emph{spin up} state in two valleys $k$($\tau=1$, $s_z=1$) and $k'$($\tau=-1$, $s_z=1$) and (b): the \emph{spin down} state in two valleys $k$($\tau=1$, $s_z=-1$) and $k'$($\tau=-1$, $s_z=-1$) and $R=5,~ 6.25,~ 7.25,~ 8.25$ for (c): the \emph{spin up} state in two valleys $k$($\tau=1$, $s_z=1$) and $k'$($\tau=1$, $s_z=1$) and (d): the \emph{spin down} state in two valleys $k$($\tau=1$, $s_z=-1$) and $k'$($\tau=-1$, $s_z=-1$).}
	\label{Fig3}
\end{figure}

To study the scattering for $E<V$ in more detail, we present in Figure \ref{Fig3} the scattering efficiency as a function of the electron energy E. In figures \ref{Fig3}(a) and  \ref{Fig3}(b), We consider dots with small radii $R = 0.01,~0.02,~ 0.03,~ 0.05$, for the \emph{spin up} state in the two valleys $k$($\tau=1$) and $k'$($\tau=1$) and the \emph{spin down} state in two valleys $k$($\tau=1$) and $k'$($\tau=-1$).

In Figure \ref{Fig3}(a), for $0\leq E \leq 0.6$ we show that $Q$ presents a maxima for the state of \emph{spin up} in the valley $k'$($\tau=-1$, $s_z=1$) with the appearance of a single peak corresponding to $E=0.55$ and a minimum for the \emph{spin up} state in the valley $k$($\tau$, $s_z=1$) without the appearance of peaks. For $E>0.6$ we show that $Q$ presents a maxima for the \emph{spin up} state in the valley $k$($\tau=1$, $s_z=1$) with the appearance of a single peak suitable for $E=0.75$ and a minima for the \emph{spin-up} state in the valley $k'$($\tau=-1$, $s_z=1$) without the appearance of peaks.

In Figure \ref{Fig3}(b), for $0\leq E \leq 0.6$ we show that $Q$ presents a maxima for the state of spin-down in the valley $k$($\tau=1$, $s_z=-1$) with the appearance of a single peak corresponds to $E=0.55$ and a minimum for the spin-up state in the valley $k'$($\tau=1$, $s_z=-1$) without the appearance of peaks. for $E>0.6$ we show $Q$ presents a maxima for the \emph{spin down} state in the valley $k'$($\tau=-1$, $s_z=-1$) with the appearance of a single peak suitable for $E=0.75$ and a minima for the \emph{spin-down} state in the valley $k$($\tau=1$, $s_z=-1$) without the appearance of peaks. However, the electron  scattering efficiency is invariant under the transformation $Q(\tau,~s_z)\rightarrow Q(-\tau,~-s_z)$.

In Figures \ref{Fig3}(c) and \ref{Fig3}(d) we present $Q$ as a function of $E$ for large values of $R=5,~ 6.25,~7.25,~ 8.25$ for the \emph{spin up} state in the two valleys $k$($\tau=1$, $s_z=1$) and $k'$($\tau=-1$, $s_z=1$) and the \emph{spin down} state in two valleys $k$($\tau=1$, $s_z=-1$) and $k'$($\tau=-1$, $s_z=-1$) respectively, we observe that $Q$ also shows large maxima for low energies. But when $E$ increases, we observe the appearance of a peak emerging with damped oscillations for both \emph{spin up} and \emph{spin down} states. These sharp peaks are due to the resonant excitation of the normal modes of the quantum dot. Consequently, in the two valleys $k$ and $k'$, the dependence of $Q$ on $s_z$ is symmetric with respect to $\pm s_z$ i.e $Q(-\tau,~s_z)=Q(\tau,~-s_z)$.

\begin{figure}[!h]\centering
	\includegraphics[scale=0.55]{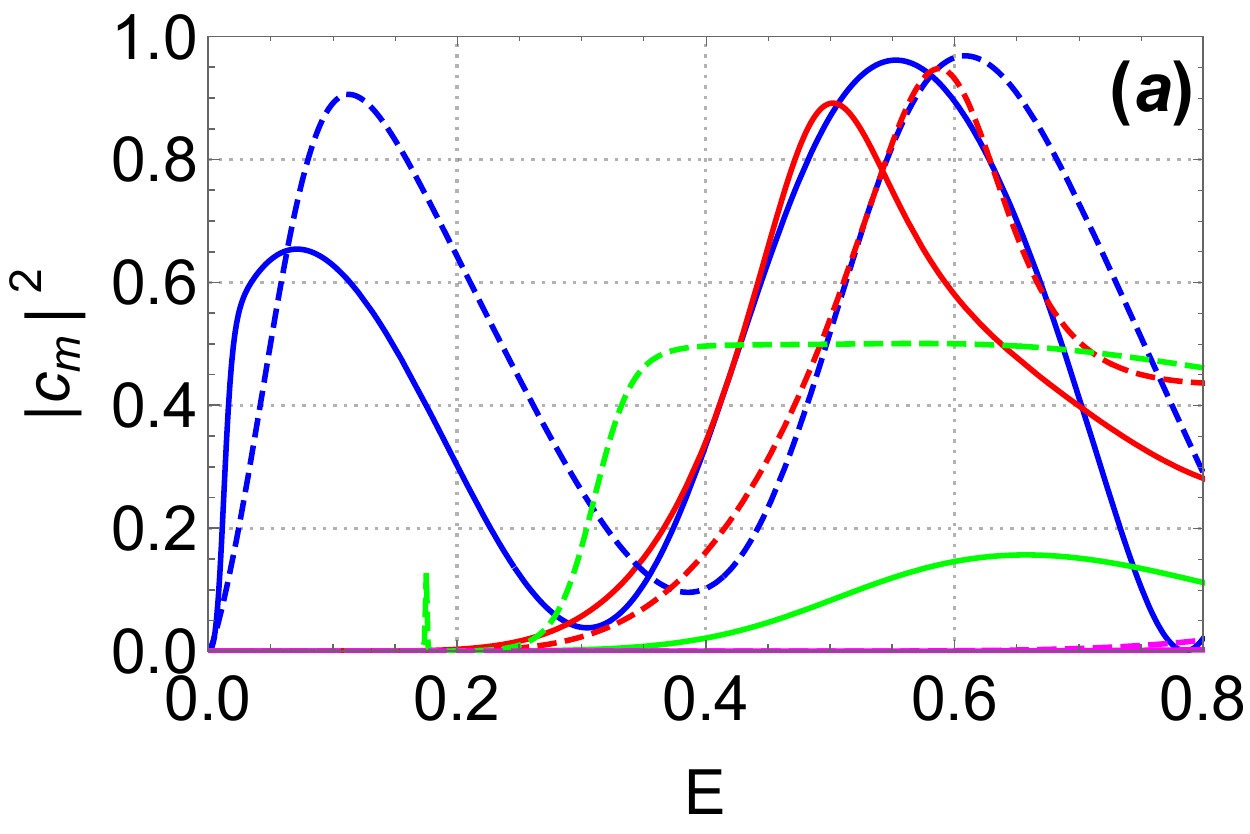}\hspace{0.5cm}\includegraphics[scale=0.55]{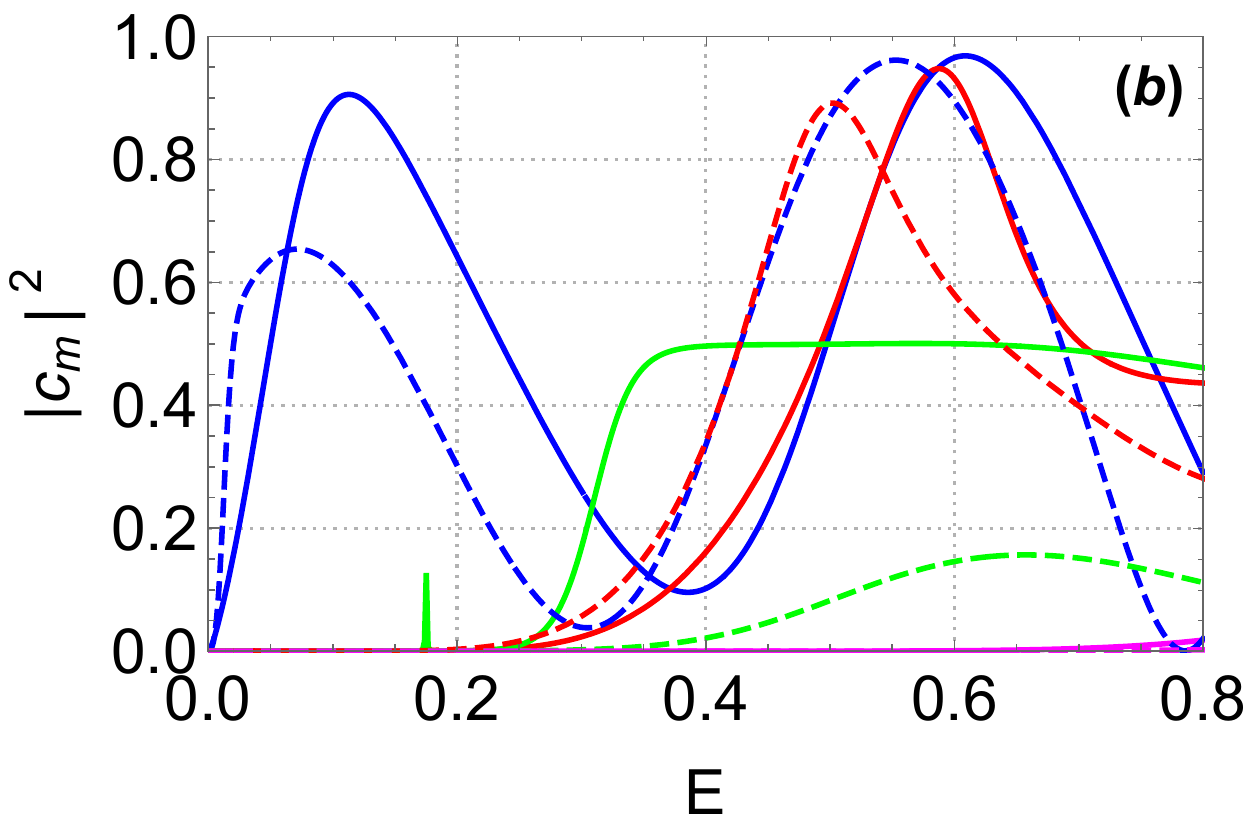} \\
	\includegraphics[scale=0.55]{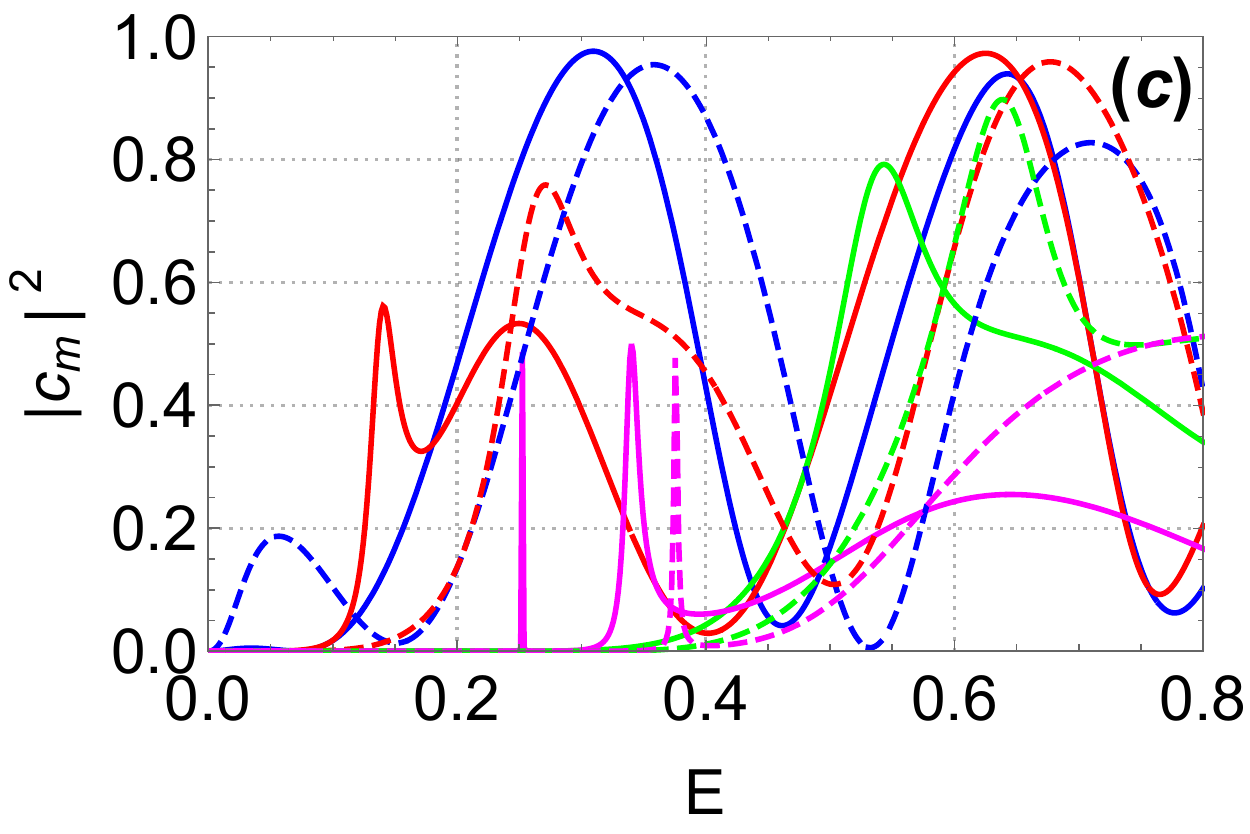}\hspace{0.5cm}\includegraphics[scale=0.55]{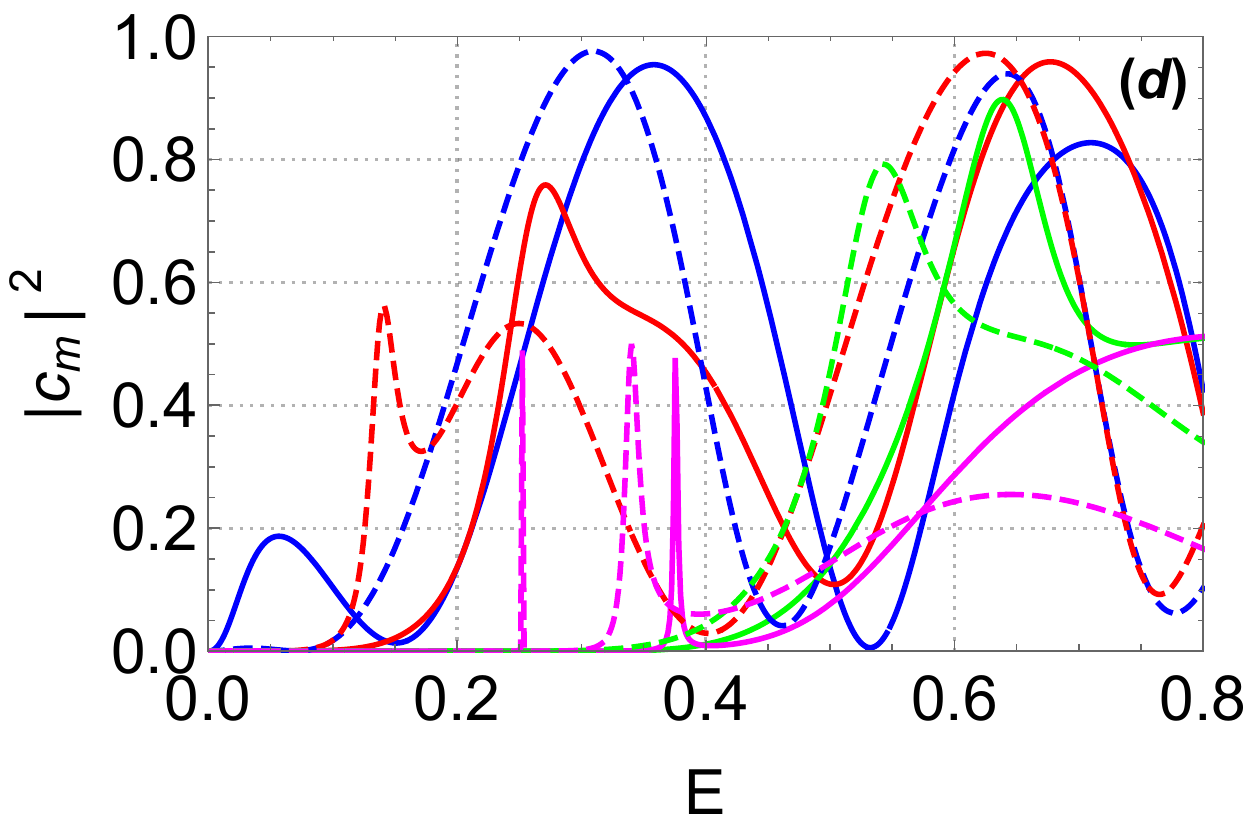}\\
	\includegraphics[scale=0.55]{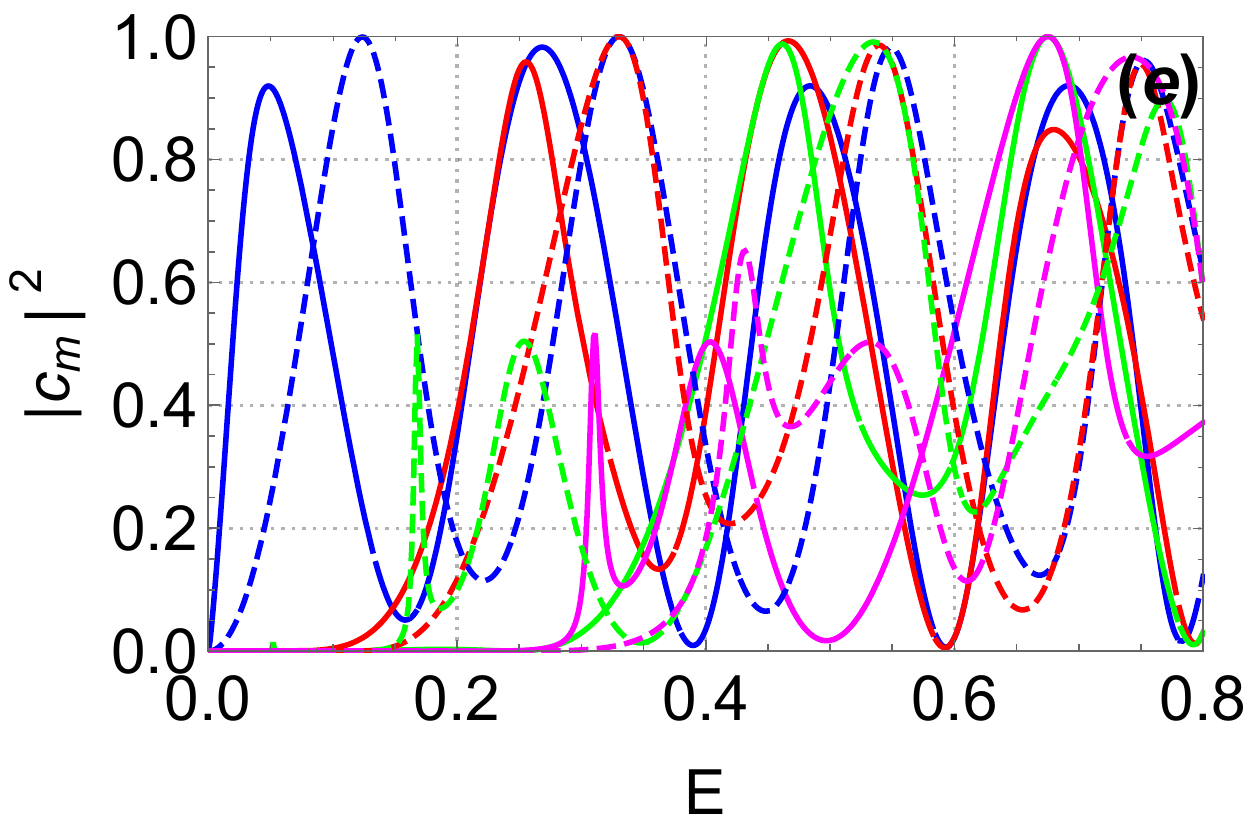}\hspace{0.5cm}\includegraphics[scale=0.55]{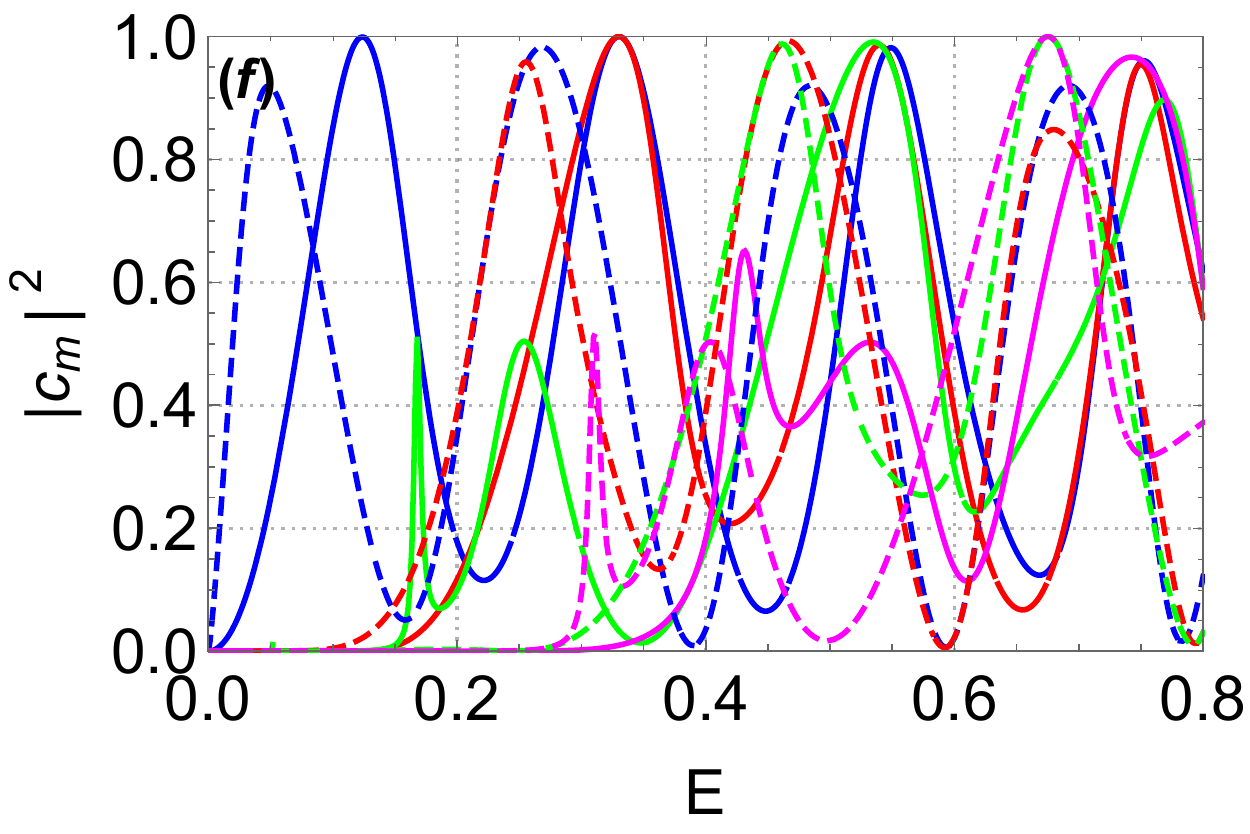} \\
	\caption{Square modulus of the scattering coefficients $|c_m|^2$, with the potential $V=1$,  for $m=0$ (blue line), $1$ (red line), $2$ (green line), $3$ (magenta line) as a function of the energy $E$ at $V=1$ for (a): ($k$, $k'$, \emph{spin up}, $R=2$) states, (b): ($k$, $k'$, \emph{spin down}, $R=2$) states, (c): ($k$, $k'$, \emph{spin up}, $R=3$) states, (d): ($k$, $k'$, \emph{spin down}, $R=3$) states, (e): ($k$, $k'$, \emph{spin up}, $R=7.75$) states and (f): ($k$, $k'$, \emph{spin down}, $R=7.75$) states. Solid line corresponds to valley $k$ and dashed line corresponds to valley $k'$ for all. }
	\label{Fig4}
\end{figure}

In Figure \ref{Fig4}, we plot the square modulus of the scattering coefficients $|c_m|^2$ for $m=0,~1,~2,~3$ as a function of the energy $E$, for the \emph{spin up} state ($s_z=1$) in the two valleys $k$($\tau=1$) and $k'$($\tau=1$) and the \emph{spin-down} state in the two valleys $k$($\tau=1$) and $k'$($\tau=1$) for different size of the dot radius : \ref{Fig4}(a) and \ref{Fig4}(b): $R=2$, \ref{Fig4}(c) and \ref{Fig4}(d): $R=3$, \ref{Fig4}(e) and \ref{Fig4}(f): $R=4$ with  in all panels $V=1$. We observe that for zero energy or close to zero energy all scattering coefficients are zero except the case corresponding to $m=0$. Also, by increasing the energy the scattering coefficients $|c_m|^2$ tend to oscillatory behavior \cite{Zheng19}. We can see that the application of the spin-orbit interaction leads to an increase in the number of oscillations. Moreover, we note the some values of the energy, $|c_m|^2$ present sharp peaks. Furthermore, these resonances of normal modes of the dot lead to the sharp peaks already observed for the scattering efficiency $Q$ as a function of the energy (Figure \ref{Fig3}). These results show that the term spin-orbit interaction of equation \eqref{eq1} requires symmetry $|c_m|^2(-\tau,~s_z)=|c_m|^2(\tau,~-s_z)$.

\begin{figure}[!h]\centering
	\includegraphics[scale=0.5]{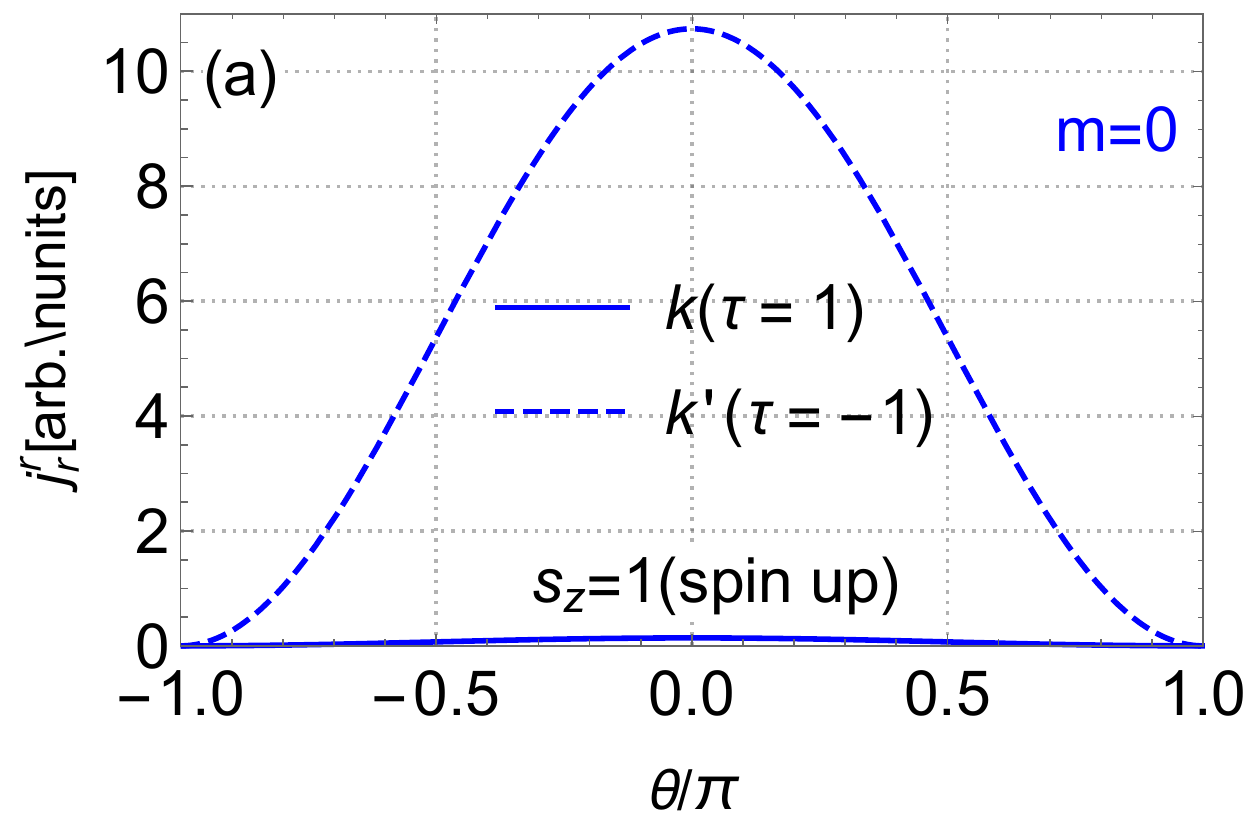}\hspace{0.5cm}\includegraphics[scale=0.5]{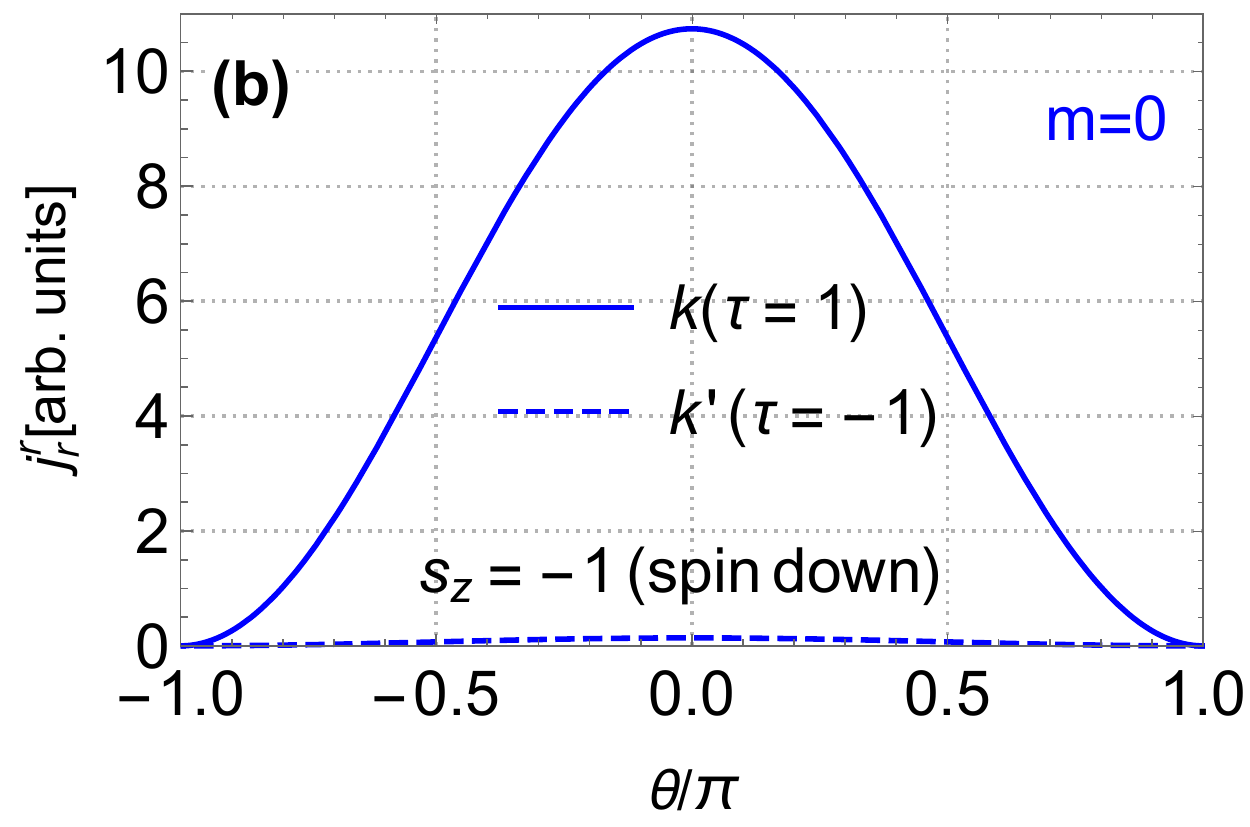}\\
	\includegraphics[scale=0.5]{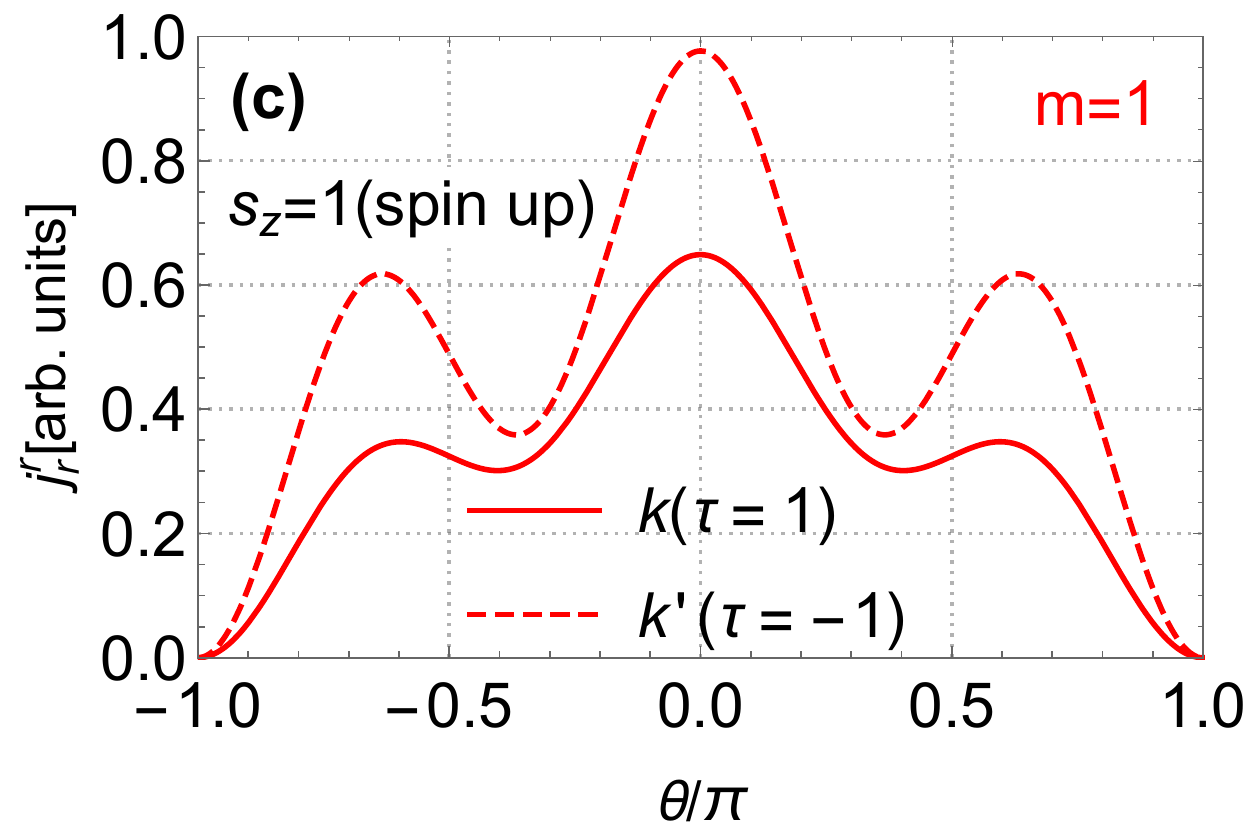}\hspace{0.5cm}\includegraphics[scale=0.5]{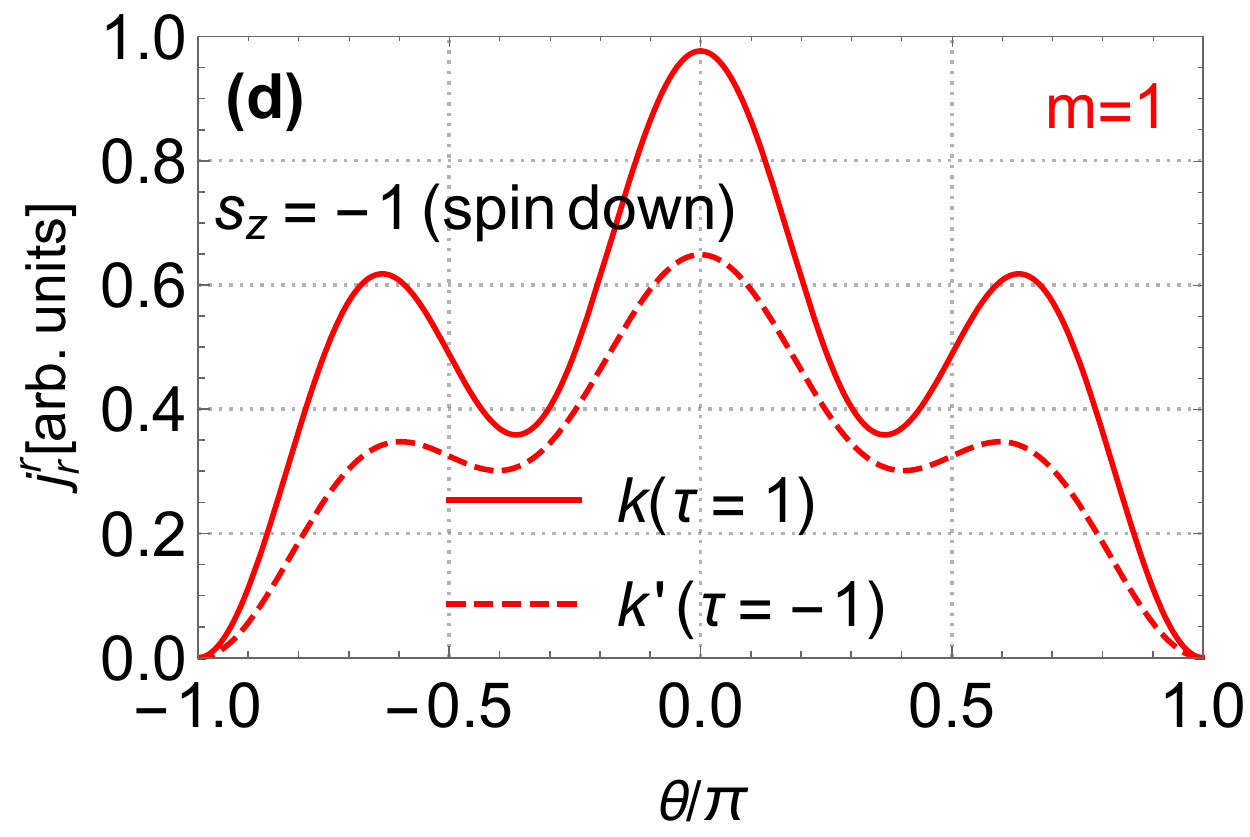}\\
	\includegraphics[scale=0.5]{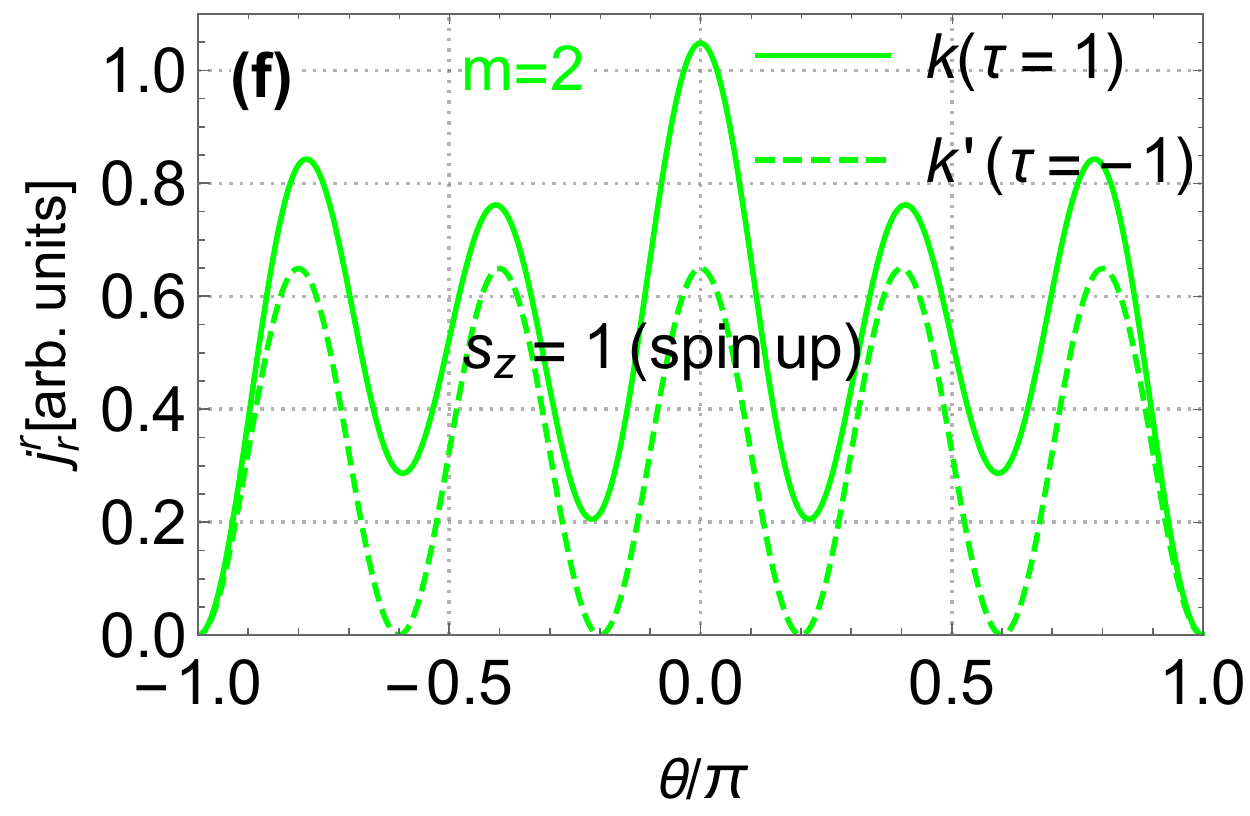}\hspace{0.5cm}\includegraphics[scale=0.5]{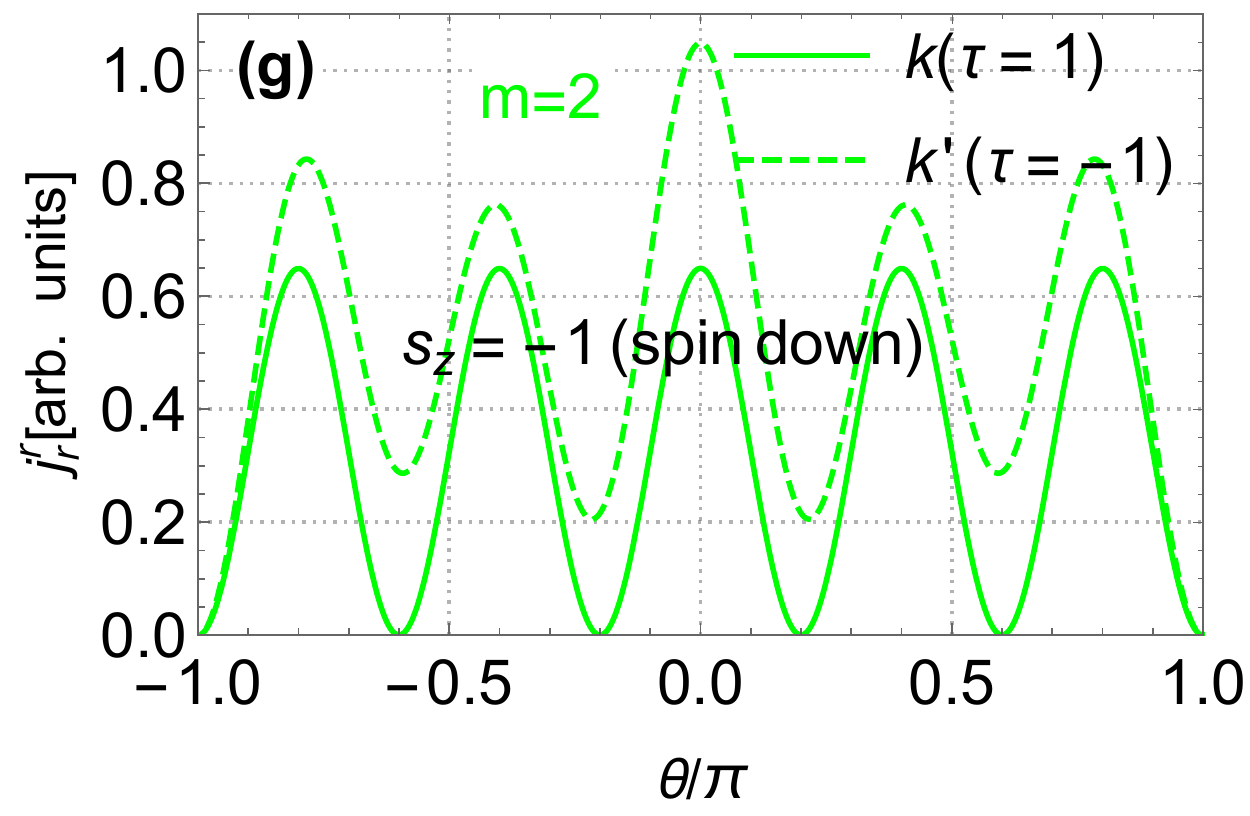}\\
	\includegraphics[scale=0.5]{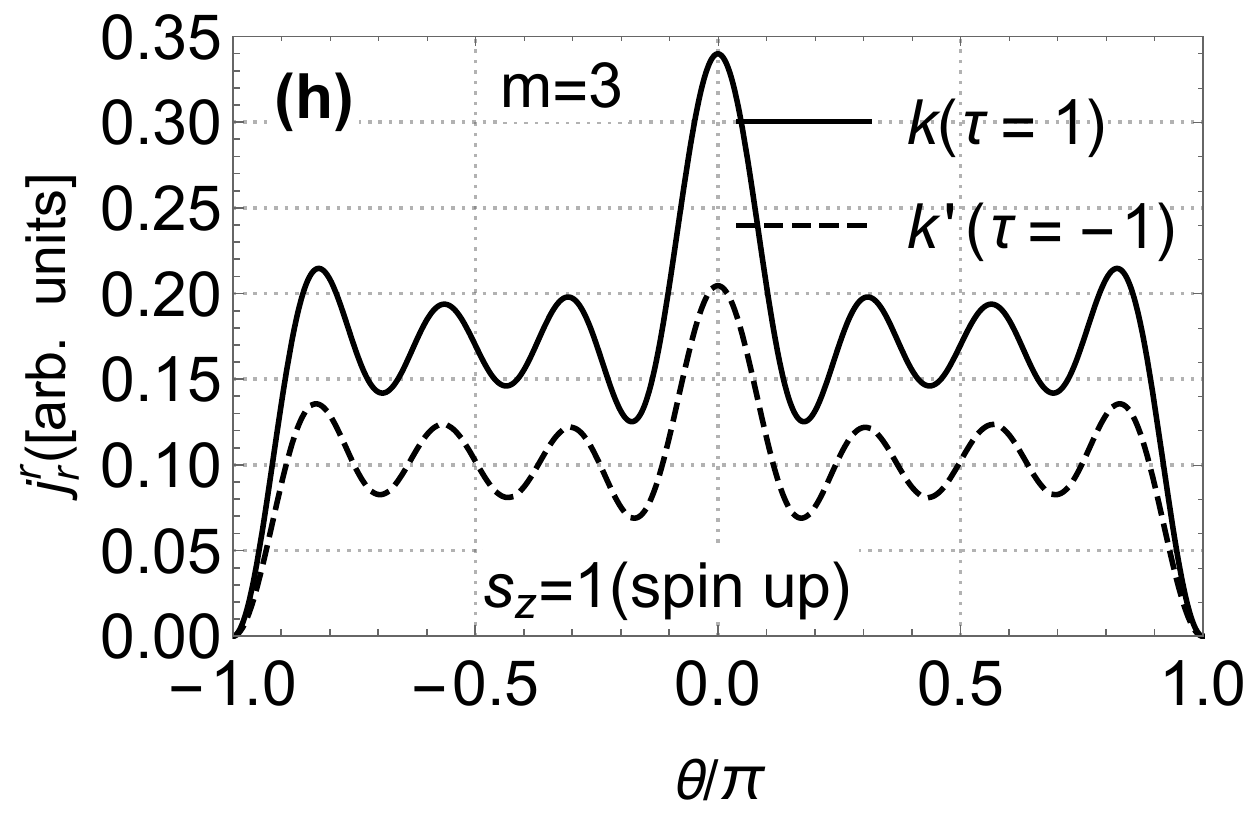}\hspace{0.5cm}\includegraphics[scale=0.5]{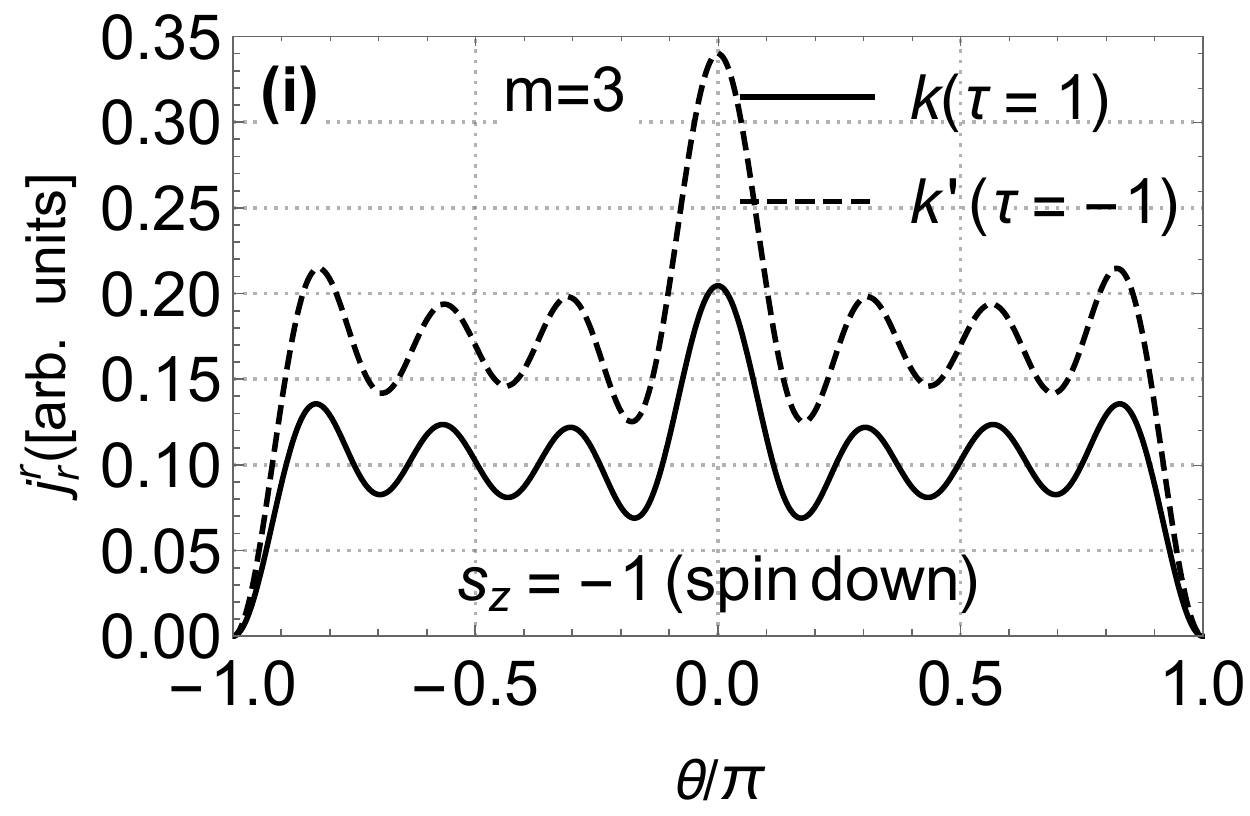}
	\caption{Radial component of the far-field scattered current $j^r_r$ as a function of the angle $\theta$ for (a): ($k$, $k'$, \emph{spin up}, $m=0$) and (b): ($k$, $k'$, \emph{spin down}, $m=0$) states, (c): ($k$, $k'$, \emph{spin up}, $m=1$) and (d): ($k$, $k'$, \emph{spin down}, $m=1$) states, (e): ($k$, $k'$, spin up, $m=2$)and (f): ($k$, $k'$, spin down, $m=2$) states, (g): ($k$, $k'$, \emph{spin up}, $m=3$) and (h): ($k$, $k'$, \emph{spin down}, $m=3$) states with $R=7.75$, $V=1$ and $E=0.0704$.}
	\label{Fig5}
\end{figure}

In Figure \ref{Fig5}, we plot the angular characteristic of the reflected radial component $j^{r}_{r}$ as a function of $\theta$ for the \emph{spin up} and \emph{spin down} states. We show that $j^{r}_{r}$ presents a maxima for $\theta=0$ and a minima for $\theta=\pm\pi$. Moreover, for the mode $c_0$ (Figures \ref{Fig5}(a) and \ref{Fig5}(b)) only forward scattering is favored. While for higher modes more preferred scattering directions emerge. There by, for $m=1$ (Figures \ref{Fig5} (c) and \ref{Fig5}(d)) three preferred scattering directions.

However, for $m=2$ (Figures \ref{Fig5} (e) and \ref{Fig5}(f)) five preferred scattering directions and for $m=3$ (Figures \ref{Fig5} (g) and \ref{Fig5}(h)) seven preferred scattering directions. In general, each mode has $2m+1$ preferred scattering direction observable but with different amplitude \cite{Heinisch13}, however the mode ($m=0$) presents a greater amplitude than the higher modes ($m>0$). Resonant scattering by only one of the normal modes is also reflected by the electron density profile near the dot. Consequently, in both the up and down states, the dependence of $j^{r}_{r}$ on $\tau$ is symmetric with reference to (regarding) $\pm\tau$ and $\pm s_z$, i.e. $j^{r}_{r}(-\tau,s_z)=j^{r}_{r}(\tau,-s_z)$.
\begin{figure}[!h]\centering
	\includegraphics[scale=0.3]{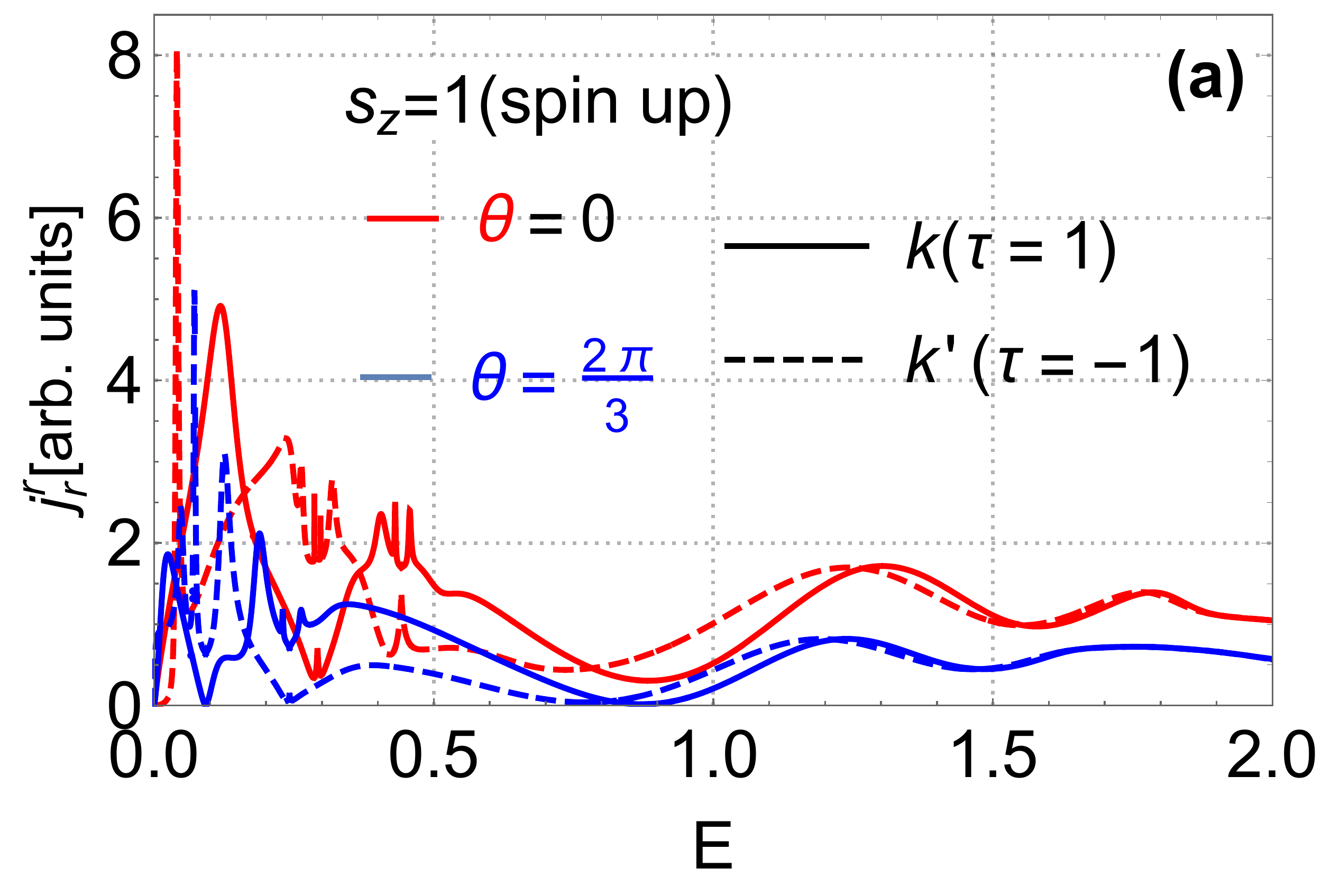}\hspace{0.5cm}\includegraphics[scale=0.3]{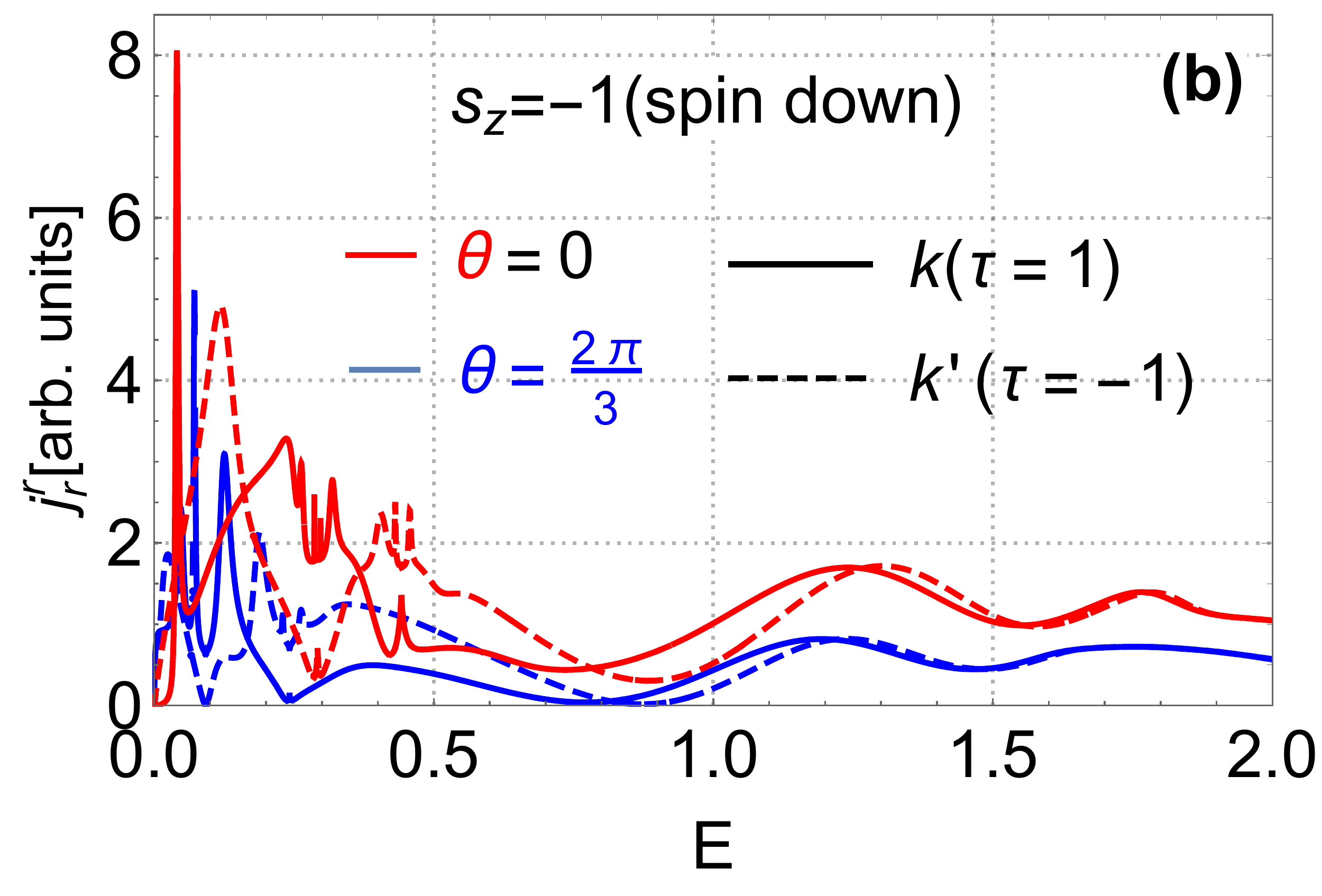}
	\caption{The radial component of the far-field scattered current $j_r^r$ as a function of the incident energy $E$ for the angles $\theta=2\pi/3$ (blue line) and $\theta=0$ (red line) with $V=1$ and $R=2$. (a): ($k$, $k'$, \emph{spin up}) and (b): ($k$, $k'$, \emph{spin down}) states.}
	\label{fig6}
\end{figure}

In figure \ref{fig6} we present the radial component of the far-field scattered current $j_r^{r}$ as a function of the incident energy for the states (a): ($k$($\tau=1$), $k’$($\tau=-1$), \emph{spin up}) and (b): ($k$($\tau=1$), $k'$($\tau=-1$), \emph{spin down}) with $V=1$ and $R=4$. In all $\theta=0$ (red line) and $\theta=2\pi/3$ (blue line) \cite{Heinisch13}. In Figure \ref{fig6}(a) we show that when $E\rightarrow 0$, $j_r^{r}$ for the two values of $\theta$, when $E$ increases to the value $0.5$, we observe the appearance of peaks of resonances with a maximum peak for $(\theta,~\tau,~s_z)=(0,~-1,~1)$. While in the regime $05<E<1.5$, $j_r^{r}$ shows an oscillatory behavior, moreover, in the regime $E\geq 1.5$, $j_r^{r}$ presents a damped oscillatory behavior with the symmetry $j_r^{r}(\theta,~\tau,~s_z)=j_r^{r}(\theta,~-\tau,~s_z)$. In Figure \ref{fig6}(b) we show that the behavior of $j_r^{r}$ is similar to that of Figure \ref{fig6}(a) when $\tau \rightarrow -\tau$ and $s_z \rightarrow -s_z$, in an equivalent way to write $j_r^{r}(\theta,~-\tau,~s_z)=j_r^{r}(\theta,~\tau,~-s_z)$.

The resonant scattering of a single mode is also reflected by the electron density profile near the dot. Inside the quantum dot, the density is given by
\begin{equation}
	\psi_t^{\dagger}\psi_t=|\beta_m|^2\left[\left(j_m(\gamma r)^2+(\mu\, j_{m+1}(\gamma r)\right)^2\right].
\end{equation}
\begin{figure}[!h]\centering
	\includegraphics[scale=0.45]{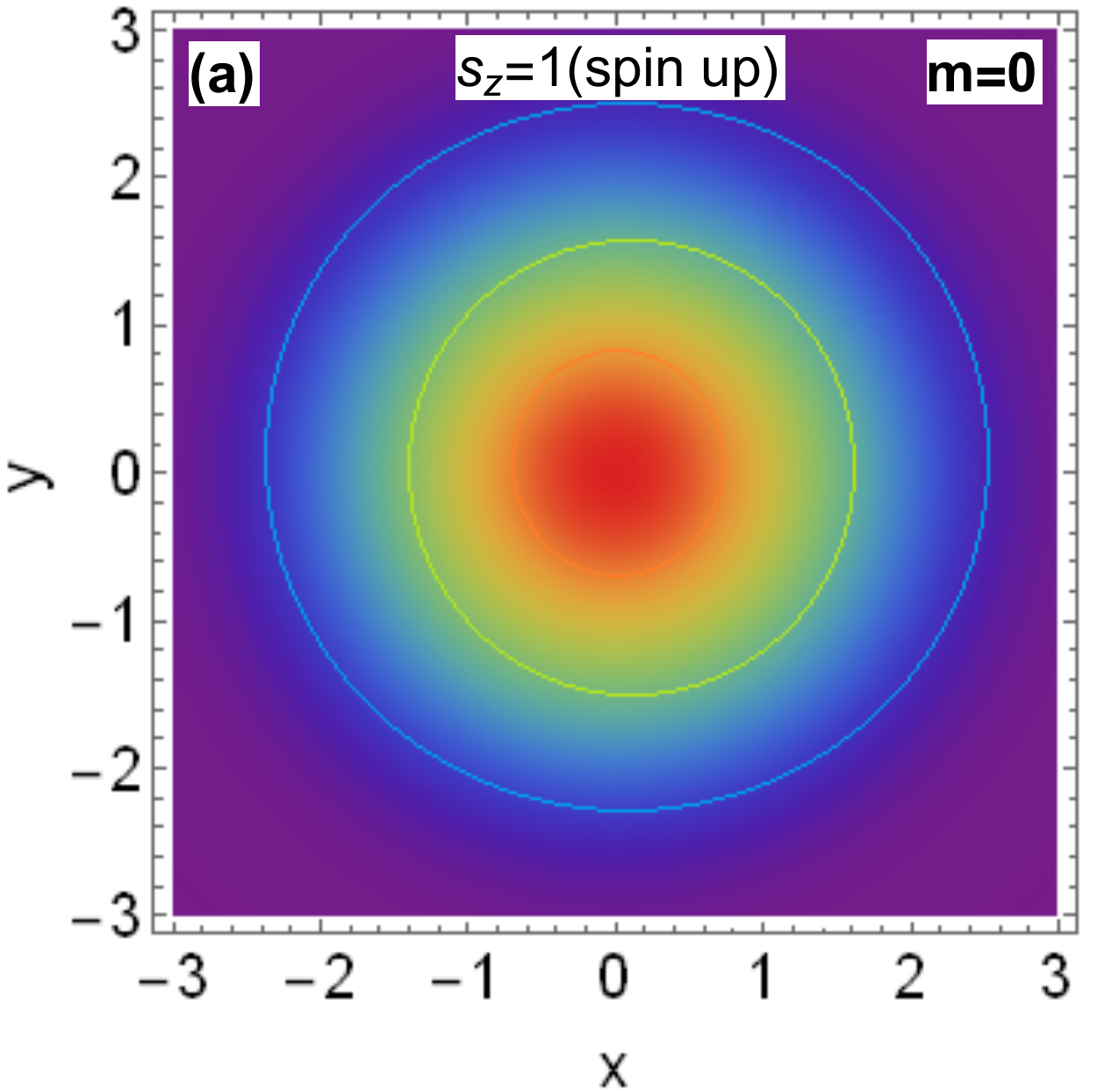}\hspace{1cm}\includegraphics[scale=0.45]{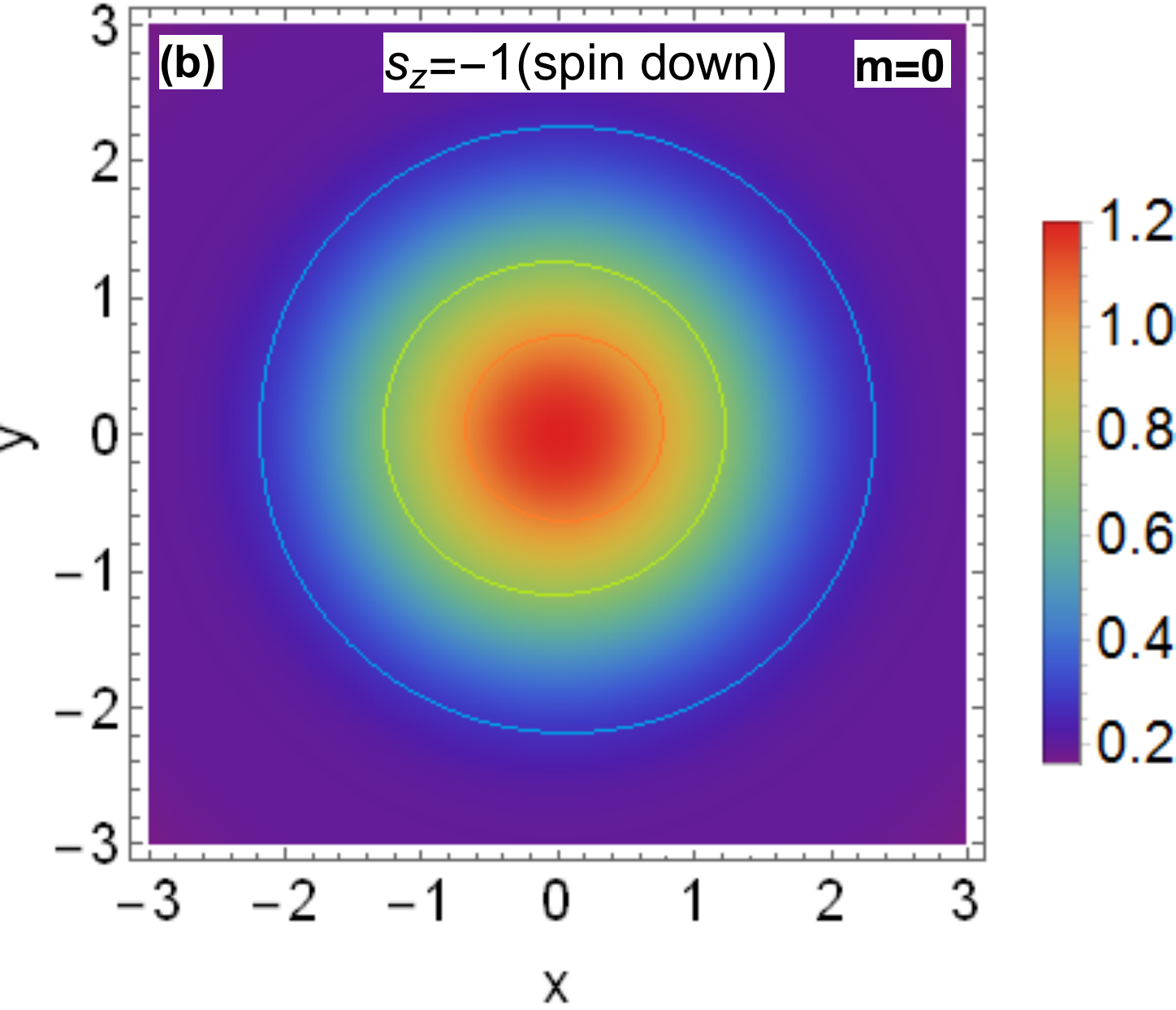}\\
	\includegraphics[scale=0.45]{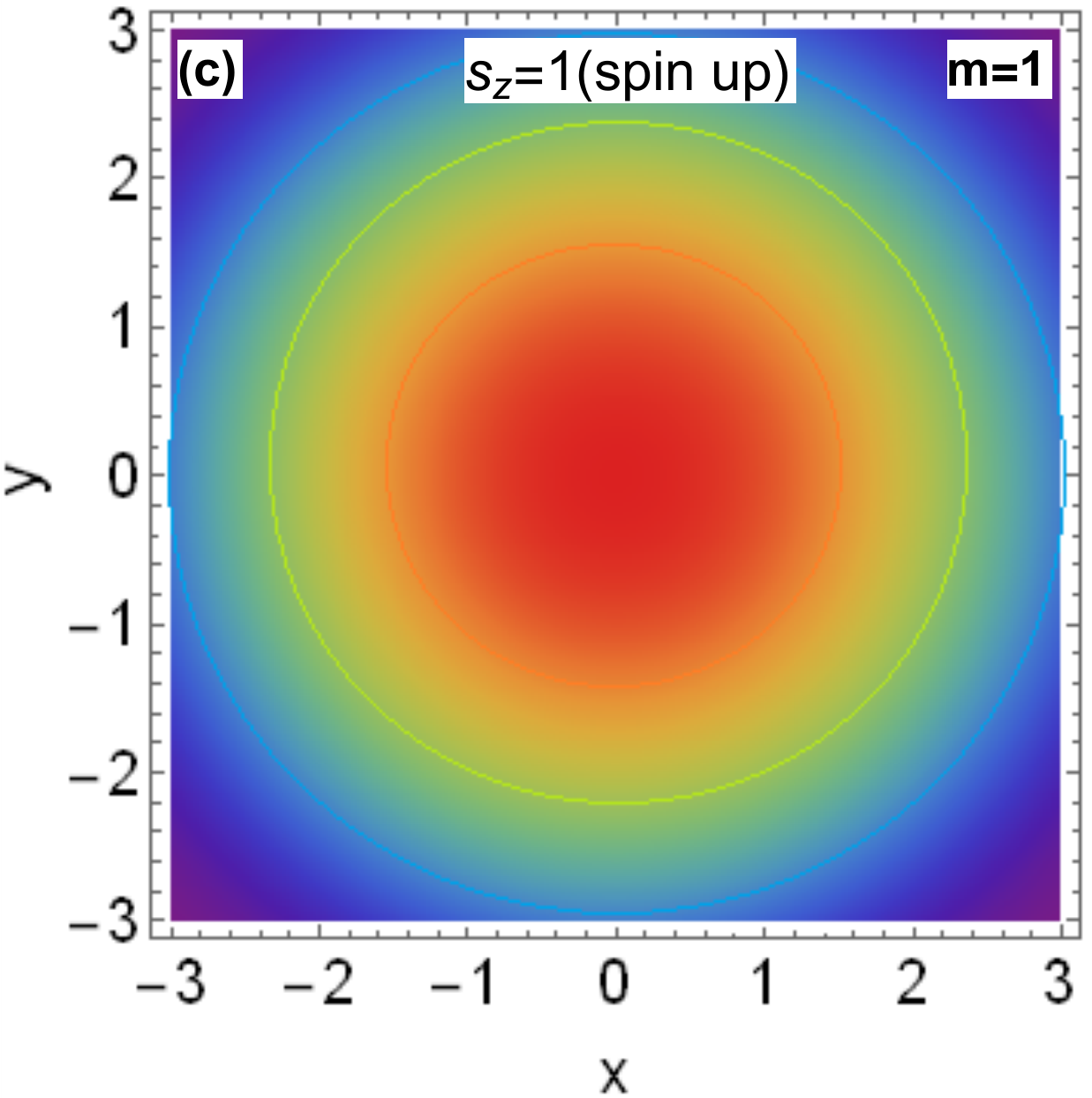}\hspace{1cm}\includegraphics[scale=0.45]{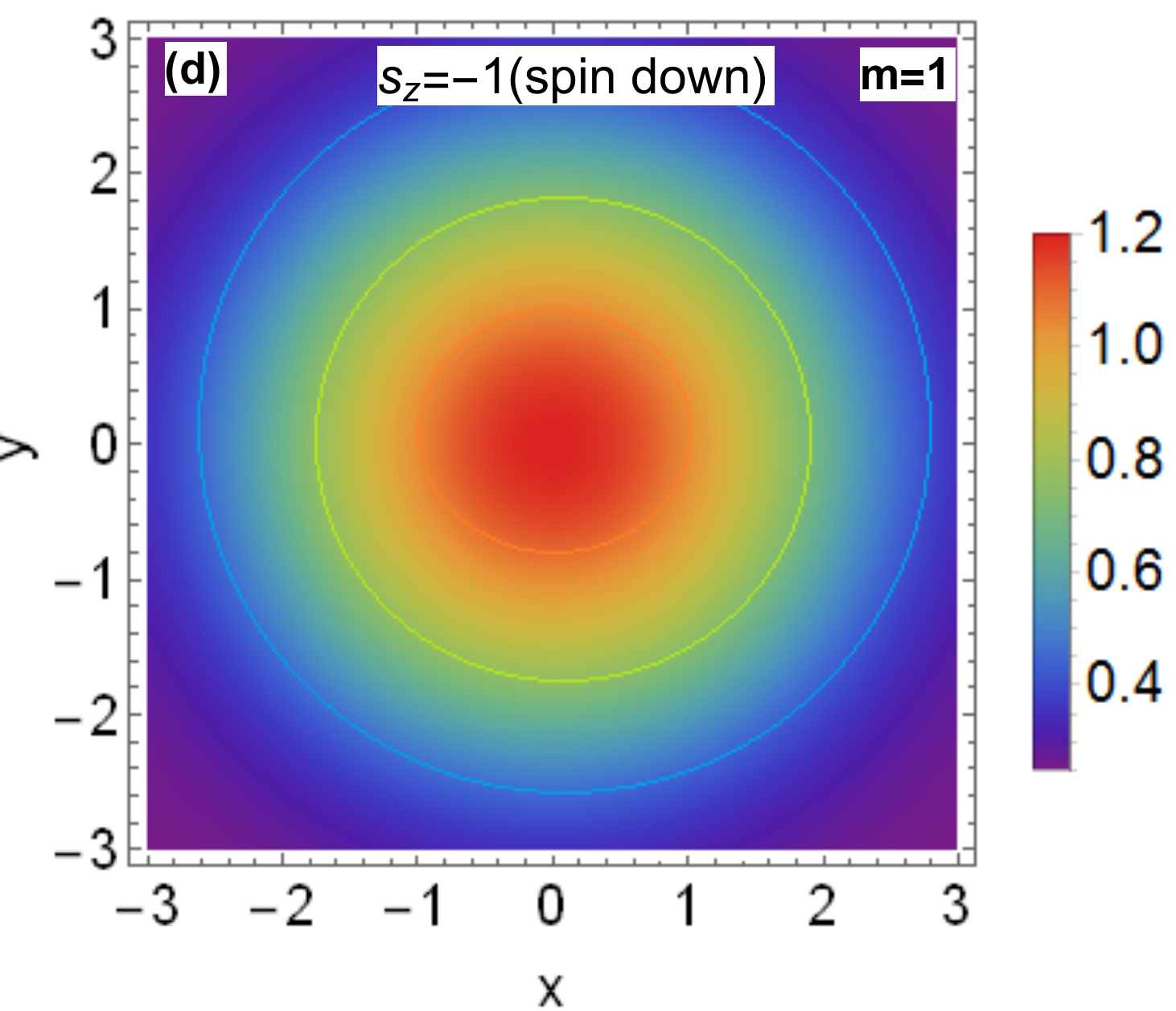}\\
	\includegraphics[scale=0.45]{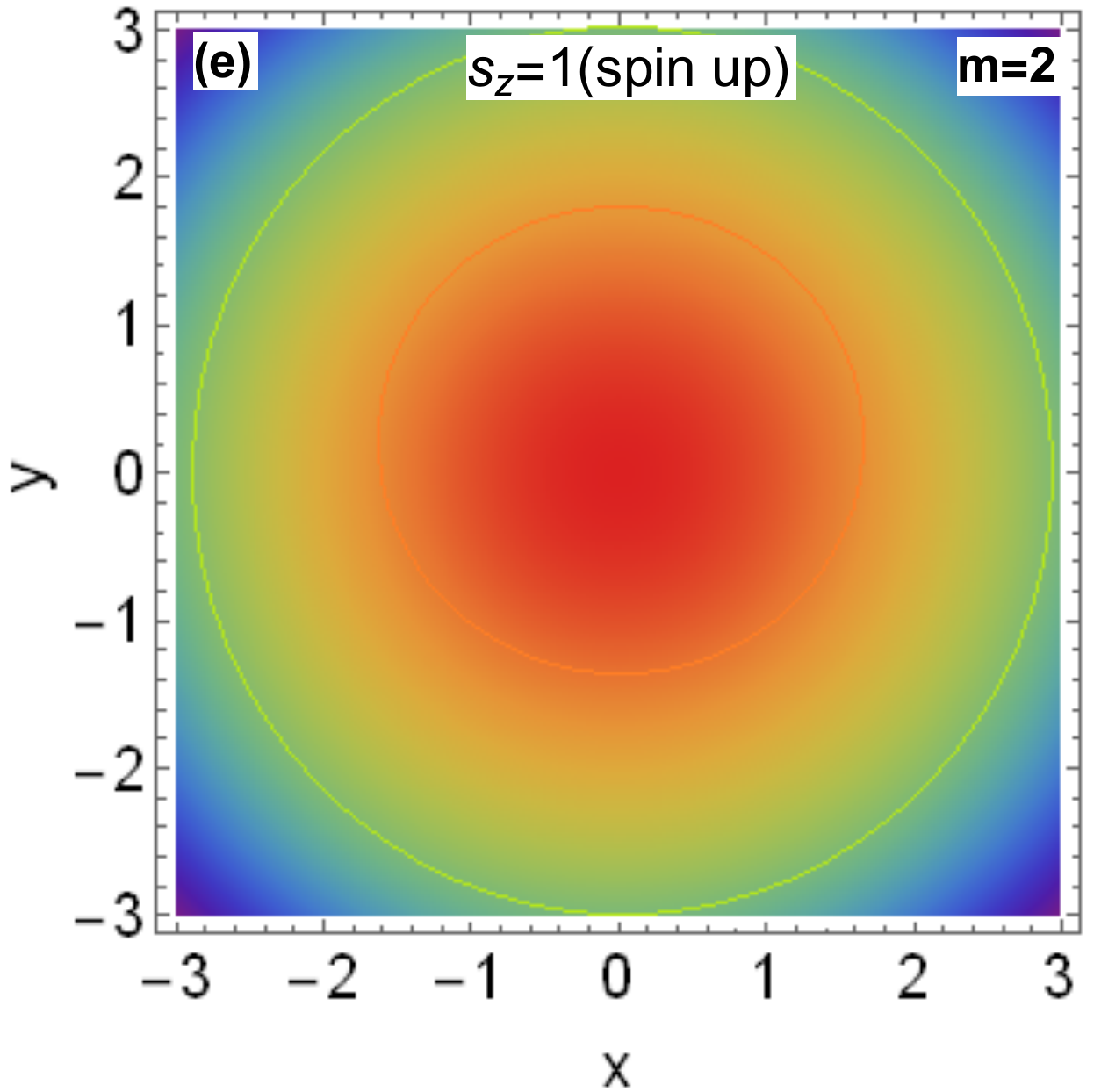}\hspace{1cm}\includegraphics[scale=0.45]{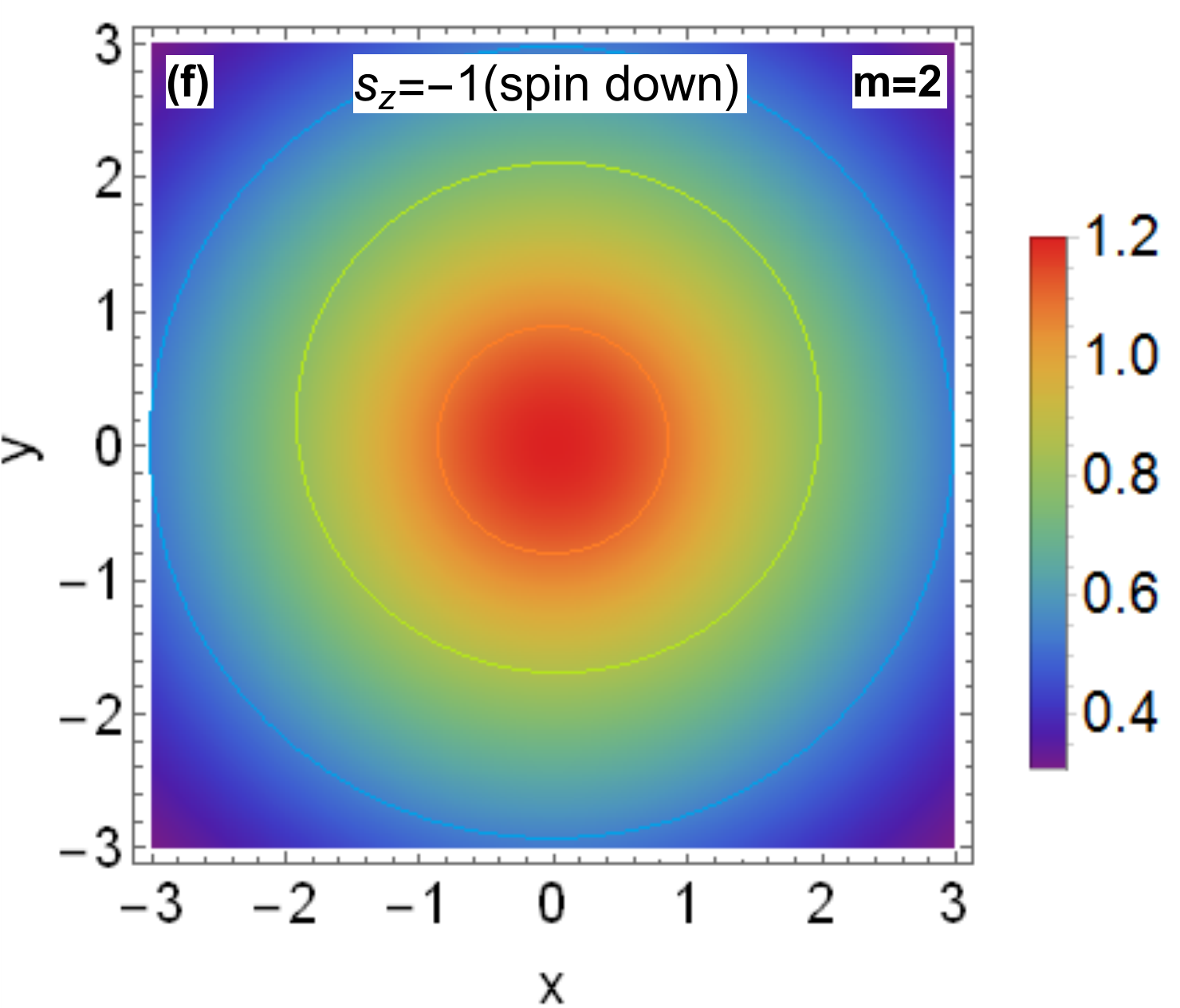}\\
	\caption{Spatial density profile $\psi_t^{\dagger}\psi_t$ in the vicinity of the quantum dot for (a): ($k$, $k'$, \emph{spin up}, $m=0$) and (b): ($k$, $k'$, \emph{spin down}, $m=0$) states (c): ($k$, $k'$, \emph{spin up}, $m=1$) and (d): ($k$, $k'$, \emph{spin down}, $m=1$) states (c): ($k$, $k'$, \emph{spin up}, $m=2$) and (d): ($k$, $k'$, \emph{spin down}, $m=2$) states for $R=3$, $E=0.078$ and $V=1$.}
	\label{Fig7}
\end{figure}

In figure \ref{Fig7}, we plot the spatial density $\psi_t^{\dagger}\psi_t$ in the quantum dot for the modes $ \beta_0 $, $ \beta_1 $, $ \beta_2 $ and $ \beta_3 $. The modes $ \beta_m $ has a maximum electron density in the center of the quantum dot and when the size of quantum dot increases the electron density decreases for both \emph{spin up}  and \emph{spin down} states. By gradually increasing the angular monument m, we also show that the electron density becomes important. In addition, the \emph{spin up}  electron density  and the \emph{spin down}  electron density  are not similar, more importantly, the electron density inside the dot is greatly increased, which is a sign of temporary trapping from particles to scattering resonances.
\section{Conclusion}\label{Conclsion}
We have studied the scattering of a plane Dirac electron wave on a circular quantum dot defined electrostatically in the monolayer of $\mathrm{MoS_2}$. We used the boundary conditions at the edges of the quantum dot to determine the scattering coefficients $\alpha_m$ and $\beta_m$ which describe the characteristics of our systems. The scattering efficiency, the square modulus of the scattering coefficient, and the radial component of the current density were calculated. The scattering of a plane Dirac electron wave has been studied in two energy regimes of the incident electron $E<V$, and $E\geq V$.

For the regime $E<V$, low energy of the incident electron, $Q$ presents a damped oscillatory behavior with the appearance of emerging peaks due to the excitation of the normal modes of the dot appear, small values of $E$ correspond to large amplitudes of $Q$. For the other regime $E\geq V$, we have shown that $Q$ has an oscillatory behavior. On the other hand, we observed a remarkable valley symmetry $Q(-\tau,~s_z)=Q(\tau,~-s_z)$.

To identify the resonances, we have studied  the energy dependence of the square module of the scattering coefficients $|c_m|^2$  it was found that near $E\rightarrow 0$, only the lowest scattering coefficient is non-null but with the increase of $E$ the remaining coefficients started to show some contributions. For larger $E$, $|c_m|^2$ tend to have oscillatory behavior, but for a not too large $E$ we have seen that the successive appearance of modes is interspersed with sudden and sharp peaks of different $|c_m|^2$. Regarding the angular characteristic of the reflected radial component, we have found that each mode has ($2m+1$) preferred directions of scattering observable with different amplitudes. Moreover, we have shown that the density of electrons inside the quantum dot are considerably increased, which is a sign of temporary trapping of electrons during the scattering resonances.

\end{document}